\documentclass{siamltex}
\usepackage{amsfonts}
\usepackage{amssymb,amsmath}
\usepackage{fullpage}
\usepackage{ucs}
\usepackage[utf8x]{inputenc}
\usepackage{amsmath}
\usepackage{amsfonts}
\usepackage{amssymb}
\usepackage{graphicx}
\usepackage{wrapfig}
\usepackage{color}
\usepackage{calrsfs}
\usepackage{float}

\newcommand{\dsp}{\displaystyle}
\newcommand{\CC}{{\rm C\!C}}
\newcommand{\by}{\mathbf{y}}

\usepackage{soul}
\usepackage{hyperref}

\begin{document}

\title{Robust seismic velocity change estimation using ambient noise recordings}

\author{E. Daskalakis \footnotemark[1], C. P. Evangelidis\footnotemark[2], J. Garnier \footnotemark[3], N. S. Melis \footnotemark[4], G. Papanicolaou \footnotemark[5] and C. Tsogka \footnotemark[6]}
\maketitle

\renewcommand{\thefootnote}{\fnsymbol{footnote}}

\footnotetext[1]{Mathematics and Applied Mathematics,
  University of Crete and IACM/FORTH, GR-71409 Heraklion,
  Greece. (edaskala@iacm.forth.gr)}
\footnotetext[2]{Institute of Geodynamics, National Observatory of Athens, Athens, Greece. 
    (cevan@noa.gr)}
\footnotetext[3]{Laboratoire de Probabilit\'es et Mod\`eles Al\'eatoires
\& Laboratoire Jacques-Louis Lions,
Universit{\'e} Paris VII,
75205 Paris Cedex 13,
France. (garnier@math.univ-paris-diderot.fr)}
\footnotetext[4]{Institute of Geodynamics, National Observatory of Athens, Athens, Greece. 
    (nmelis@noa.gr)}
\footnotetext[5]{Mathematics Department,
Stanford University,
Stanford, CA 94305.
(papanicolaou@stanford.edu)}
\footnotetext[6]{Mathematics and Applied Mathematics,
  University of Crete and IACM/FORTH, GR-71409 Heraklion,
  Greece. (tsogka@uoc.gr)}

\begin{abstract}

We consider the problem of seismic velocity change estimation using ambient noise recordings. 
Motivated by \cite{clayton} we study how the velocity change estimation is affected by seasonal fluctuations in the noise sources. 
More precisely, we consider a numerical model and introduce spatio-temporal seasonal fluctuations in 
the noise sources.  We show that indeed, as pointed out in \cite{clayton}, the stretching method is affected by these fluctuations 
and produces misleading apparent velocity variations which reduce dramatically the signal to noise ratio of the method.
We also show that these apparent velocity variations
can be eliminated by an adequate normalization of the cross-correlation functions. Theoretically we expect our approach to work as long as the seasonal fluctuations in the noise sources are uniform, an assumption which holds for closely located seismic stations. We illustrate with numerical simulations and real measurements that the proposed normalization significantly improves the accuracy of the velocity change estimation.
\end{abstract}


\section{Introduction}
We are interested in monitoring volcanic edifices for temporal changes of the velocity of the seismic waves. When magma pressure 
increases inside a volcano, the added pressure results into the inflation of the volcano, and small cracks around the magma chamber 
will decrease the velocity of seismic waves. That small decrease in velocity can be detected using travel-time tomography of seismic 
waves and up until very recently only the seismic waves generated by natural events like earthquakes could be used 
\cite{poupinet1984, Ratd1995, Gret2005}. There are however limitations that make the use of such seismic events not suitable 
for monitoring, like the repeat rate or the unknown source position. In recent years ambient seismic noise recordings have been 
successfully used instead of seismic events \cite{towards2008, duputel2009}.

The idea that has been exploited is that information about the Green's function or the travel-time between two seismic stations can 
be obtained from cross-correlations (\CC) of ambient noise recordings \cite{curtis06, schuster09, cross, wap10a, wap10b}.  
A number of passive imaging studies based on this idea are now used in volcano monitoring \cite{towards2008, duputel2009}, 
in seismic faults studies \cite{brenguier2008b,Acarel2014} and more generally in studying the structure of the crust 
\cite{Acarel2014, lunar2010}.  In the case of volcano monitoring, there is a large number of studies concerning 
Piton de la Fournaise, which is a shield volcano on the eastern side of Reunion island in the Indian Ocean. The goal in this setting 
is to measure relative velocity changes ($dv/v$) of surface waves which are precursors to specific events (volcanic eruptions). 
Two techniques have been used for $dv/v$ measurements, the moving window cross spectral (MWCS) method \cite{mwcs} 
and the Stretching Method (SM).

Both MWCS and SM use two waveforms, the reference and the current $\CC$ functions  which are obtained by averaging 
daily $\CC$ functions over a large, respectively a small, period of time. Changes in the velocity of the medium are estimated 
from differences in these two $\CC$ functions. In MWCS, $dv/v$ is obtained by estimating the time delays $dt_i$ in different 
time windows. The time delay estimation is performed in the frequency domain using the cross spectrum of the windowed 
wavefront segments. Then $dv/v(=-dt/t)$ is computed using a linear regression approach. SM operates in the time domain 
by solving an optimization problem which determines the stretching parameter that maximizes the correlation between 
the two waveforms.

There are some factors such as the quality and the distribution of the noise sources that can affect the temporal resolution of the 
measurements. The volcano of Piton de la Fournaise is a very well equipped area with lots of high quality stations. Moreover 
the type of the volcano (shield volcano), which is erupting very frequently, makes it an ideal example for study. That is not the case 
for most other volcanoes, especially for volcanic islands and "ring of fire volcanoes" which are poorly equipped and which erupt rarely. 
Another difficulty is that in some cases, and especially in the case we will consider in this paper, the evolution of the volcano is very 
slow and therefore long term fluctuations such as seasonal variations \cite{clayton, labasin} can hide velocity variations that 
actually correspond to volcanic activity.

In \cite{clayton} it is stated that the seasonal variations in the cross-correlations and the estimated velocity such as observed 
in \cite{labasin} are caused by seasonal variations of the amplitude spectra of the ambient noise sources. Since SM operates 
directly in the time domain it is much more likely to be affected by those seasonal variations than the MWCS method which only 
relies on the phase spectra of the cross-correlations. The stability of MWCS to spatio-temporal variations of the noise sources
is studied in \cite{colombi}. It is shown that in scattering media azimuthal variations in the intensity distribution of the noise sources 
does not affect the MWCS measurement when the coda part of the cross-correlation is used. This is because  
the anisotropy of the noise sources is mitigated by the multiple scattering of the waves with the medium inhomogeneities.

We present here a set of numerical simulations that leads to the conclusion that indeed the stretching method can produce apparent velocity variations caused by seasonal spatio-temporal fluctuations of the amplitude spectra of the noise sources. These variations are reduced 
by considering the coda part of the cross-correlations but they still persist. When the seasonal fluctuations are uniform with respect to the noise source locations, an hypothesis that is reasonable when the measurements concern recordings at the same area, the apparent velocity variations can be effectively removed by an adequate normalization (spectral whitening) of  the cross-correlated signals.  Our approach significantly improves the signal to noise ratio of the stretching method as illustrated by numerical simulations and real measurements for two volcanos. 

\section{Seasonal variations and the effectiveness of spectral whitening}
By measuring velocity variations for a long enough period using the stretching method  in \cite{labasin}, small seasonal variations were observed, which were attributed to hydrological and thermoelastic variations. In contrast,  \cite{clayton} suggests that such variations are not necessarily due to changes in the medium and could be caused by seasonal fluctuations in the amplitude spectra of the noise sources. We investigate here this question using numerically simulated data, as well as seismic noise recordings. Let us first briefly review the MWCS and the SM methods.

\subsection{Description of the moving window cross-spectral method and the stretching method}
Two methods have been predominately used for estimating velocity variations: the Stretching Method (SM)  and the Moving Window Cross-Spectral (MWCS) method \cite{mwcs}. In both methods, relative changes in the velocity of the medium are estimated by comparing two waveforms: the reference and the current cross-correlation functions which are obtained by cross-correlating the signals recorded at two different receivers over a certain period of time. The reference cross-correlation is usually the average of the daily cross-correlations over a long period of time 
of the order of a year. The current cross-correlation is a local average of the daily cross-correlation over a few days.

SM operates in the time domain and computes the stretching parameter that maximizes the correlation coefficient between the two waveforms  in a selected time window. More precisely, if $\CC_r(t)$ and $\CC_c(t)$ denote the reference and the current cross-correlation functions, then SM seeks for the stretching coefficient $\epsilon=dt/t=-dv/v$ that maximizes the following quantity,
\begin{equation}
\label{eq:10}
\dsp C(\epsilon)=\dfrac{\dsp \int_{t_1}^{t_2} \CC_{c,\epsilon}(t) \CC_r(t) dt}{\sqrt{\dsp \int_{t_1}^{t_2} (\CC_{c,\epsilon}(t))^2dt}
\dsp \sqrt{\int_{t_1}^{t_2} (\CC_r(t))^2dt}},
\end{equation}
 where $\CC_{c,\epsilon}(t)=\CC_c(t(1+\epsilon))$ is the stretched version of $\CC_c(t)$. 
 The time window $[t_1,t_2]$ is usually selected  so as to contain the coda part of the cross-correlation function and not the first arrival.

The MWCS method is described in detail in \cite{mwcs} and basically consists in computing time delays ($dt_i$) in different time windows and then estimating $dt/t$ using a linear regression model. The relative velocity change in the medium is deduced by the relationship 
$dv/v=-dt/t$.  The estimation of the time delays $dt_i$ between the reference and the current cross-correlation is performed by computing phase differences in the frequency domain.

\subsection{The numerical model} \label{sec:num}
\begin{figure}
\begin{center}
\includegraphics[width=0.5 \textwidth]{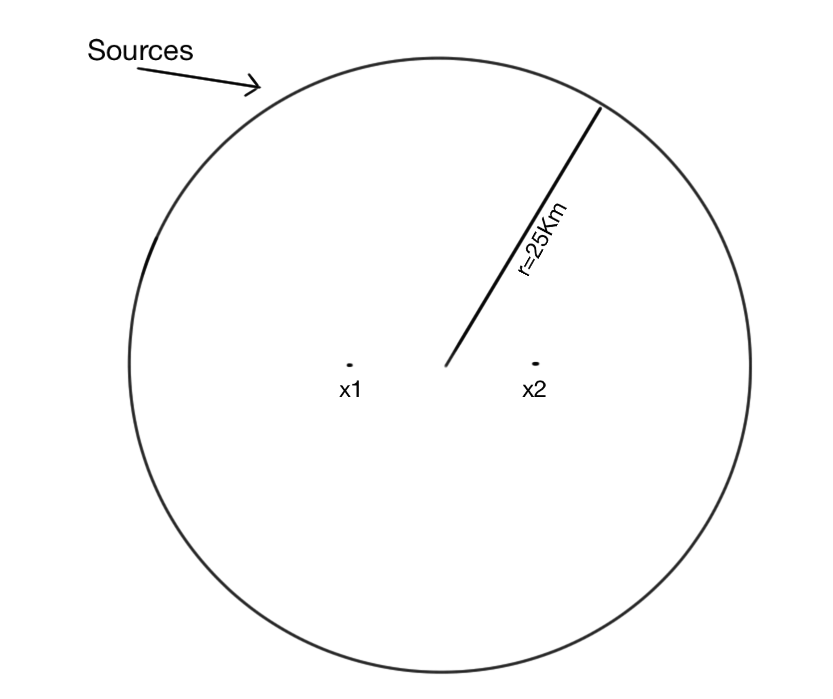}
\caption{Location of the noise sources on a circle,  ${\cal C}$, of radius $25$Km and the two receivers at $\mathbf{x}_1$ and $\mathbf{x}_2$. The distance between the two receivers is $10$Km.}
\label{fig:1}
\end{center}
\end{figure}

We carry out a set of numerical simulations that are based on a mathematical model of wave propagation. The details of the numerical model are presented 
in the appendix; here we present some basic elements and the results of those simulations. In our numerical model we consider the acoustic wave equation:
\begin{equation}
\label{eq:1}
\dfrac{1}{c(\mathbf{x})^2}\dfrac{\partial^2 u}{\partial t^2}(t,\mathbf{x})-\Delta_{\mathbf{x}}u(t,\mathbf{x})=n(t,\mathbf{x}),
\end{equation}
where $n (t,\mathbf{x})$ models the noise sources which are located on a circle,  ${\cal C}$, of radius 25km as illustrated in Figure \ref{fig:1}. We assume that the wave field is recorded at two receivers 
$\mathbf{x}_1=(-5,0)$km and $\mathbf{x}_2=(5,0)$km.\\

The solution of (\ref{eq:1}) at a given point $\mathbf{x}$ can be written as,
\begin{equation}
\label{eq:3}
u(t,\mathbf{x})=\int\int G^j(t-s,\mathbf{x},\mathbf{y})n(s,\mathbf{y})d\mathbf{y}ds,
\end{equation}
or equivalently in the frequency domain,
\begin{equation}
\label{eq:4}
\hat{u}(\omega,\mathbf{x})= \int \hat{G}^j(\omega,\mathbf{x},\mathbf{y})\hat{n}(\omega,\mathbf{y})d\mathbf{y}.
\end{equation}

Here $j$ denotes the dependence on the day, hat denotes the Fourier transform and $\hat{G}^j(\omega,\mathbf{x},\mathbf{y})$ is the Green's function.
For simplicity and easiness of the computations we consider first a homogeneous medium in which case
$\hat{G}^j(\omega,\mathbf{x},\mathbf{y})$ is given by
\begin{equation}
\label{eq:5}
\hat{G}^j(\omega,\mathbf{x},\mathbf{y})=\dfrac{1}{4\pi |\mathbf{x}-\mathbf{y}|}e^{i\frac{\omega}{c^j}|\mathbf{x}-\mathbf{y}|}.
\end{equation}
In \eqref{eq:5}, we use the 3d expression for the Green's function of the wave equation instead of the Hankel function. For our setup where the distance between the receivers is relatively large with respect to the wavelength this does not affect the results given that we are interested in the phase of the Green's function.
In \eqref{eq:5}, the velocity is allowed to change as a function of time on the scale of a day. We denote by $c^j$ the homogeneous velocity of the medium on day $j$. To illustrate the generality of our approach we also consider inhomogeneous scattering media for which the Green's function $\hat{G}^j(\omega,\mathbf{x},\mathbf{y})$ is computed by solving numerically the wave equation \eqref{eq:1} in the time domain using the code Montjoie \href{http://montjoie.gforge.inria.fr/}{(http://montjoie.gforge.inria.fr/)}.

\noindent {\bf Reference and Current cross-correlation function}
Our main tool, the daily cross-correlation function is given by
\begin{equation}
\label{eq:7}
\CC^j(\tau,\mathbf{x}_1,\mathbf{x}_2)=\frac{1}{T} \int_0^T u^j(t+\tau,\mathbf{x}_1)u^j(t,\mathbf{x}_2)dt,
\end{equation} 
with $T=24$ hours.\\
For both SM and MWCS methods, variations in the velocity are estimated by comparing two waveforms: the reference and the current cross-correlation functions. 
The reference cross-correlation will be the average of all the available daily cross-correlation functions,
\begin{equation}
\label{eq:8}
\dsp \CC_r(\tau,\mathbf{x}_1,\mathbf{x}_2)=\frac{1}{N_d} \sum_{j=1}^{N_d} \CC^j(\tau,\mathbf{x}_1,\mathbf{x}_2) ,
\end{equation}
where $N_d$ is  the total number of days,
while the current cross-correlation function that corresponds to the j-th day will be the average of 
a small number of daily cross-correlation functions around the j-th day,
\begin{equation}
\label{eq:9}
\dsp \CC_c^j(\tau,\mathbf{x}_1,\mathbf{x}_2)=\frac{1}{2s+1} \sum_{k=j-s}^{j+s} \CC^k(\tau,\mathbf{x}_1,\mathbf{x}_2).
\end{equation}
The total number of daily cross-correlations used for the current cross-correlation is $N_{ccc}=2s+1$. Usually a few days ($N_{ccc}=3$ to $10$) 
are used for the current cross-correlation while the reference one is computed for a much longer period of the order of a year \cite{mwcs}.

\begin{figure}
\begin{center}
\includegraphics[width=0.4 \textwidth]{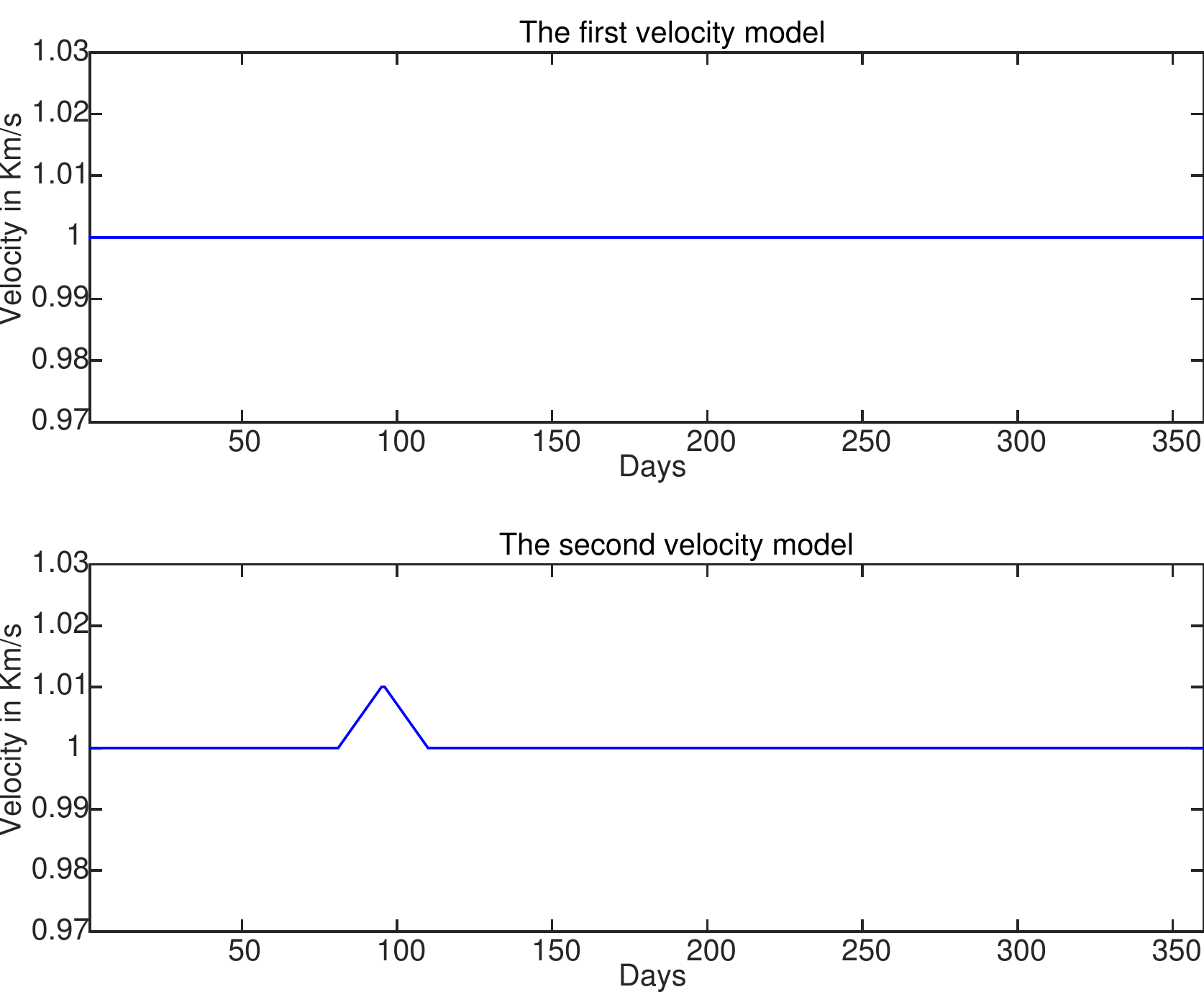}
\caption{The two velocity models. In the top plot the velocity does not change with time and is equal to $1$Km/s. In the bottom plot the velocity increases linearly between days 80 and 95 to reach the value of $1.01$Km/s and then decreases linearly with the same rate to reach its original value of $1$Km/s at day 110.}
\label{fig:2}
\end{center}
\end{figure}

\noindent {\bf Velocity Model and selected bandwidth}
We will work in the frequency bandwidth  $[0.15 - 0.65]$Hz and the total number of days is $N_d=360$ (we call this a year). 
For our simulations we consider two different velocity models, in the first case the velocity of the medium does not change with time and is equal to $1$Km/s while in the second case there is a small change in the velocity of the order of $1\%$ that takes place between days 80 to 110. The velocity increases linearly the first 15 days until it reaches the maximal value of $1.01$Km/s and then decreases linearly with the same rate to its original value of  $1$Km/s as illustrated in Figure \ref{fig:2} (bottom plot). All these numbers are realistic and very similar to the values that we have in our seismic noise recordings of the Santorini volcano considered in section \S \ref{santorini}. We have chosen the numerical set up to be similar to the experimental set up so that the numerical results are meaningful to demonstrate that the conclusions extracted from the experimental data are reliable.

\noindent {\bf Estimation of  the relative change in the velocity} 
We have implemented both SM and MWCS methods using as reference cross-correlation the average of all daily cross-correlation (360 days) and as 
current cross-correlation a $N_{ccc}=7$-day average around the day we make the measurement.

\begin{figure}
\begin{center}
\includegraphics[width=0.5 \textwidth]{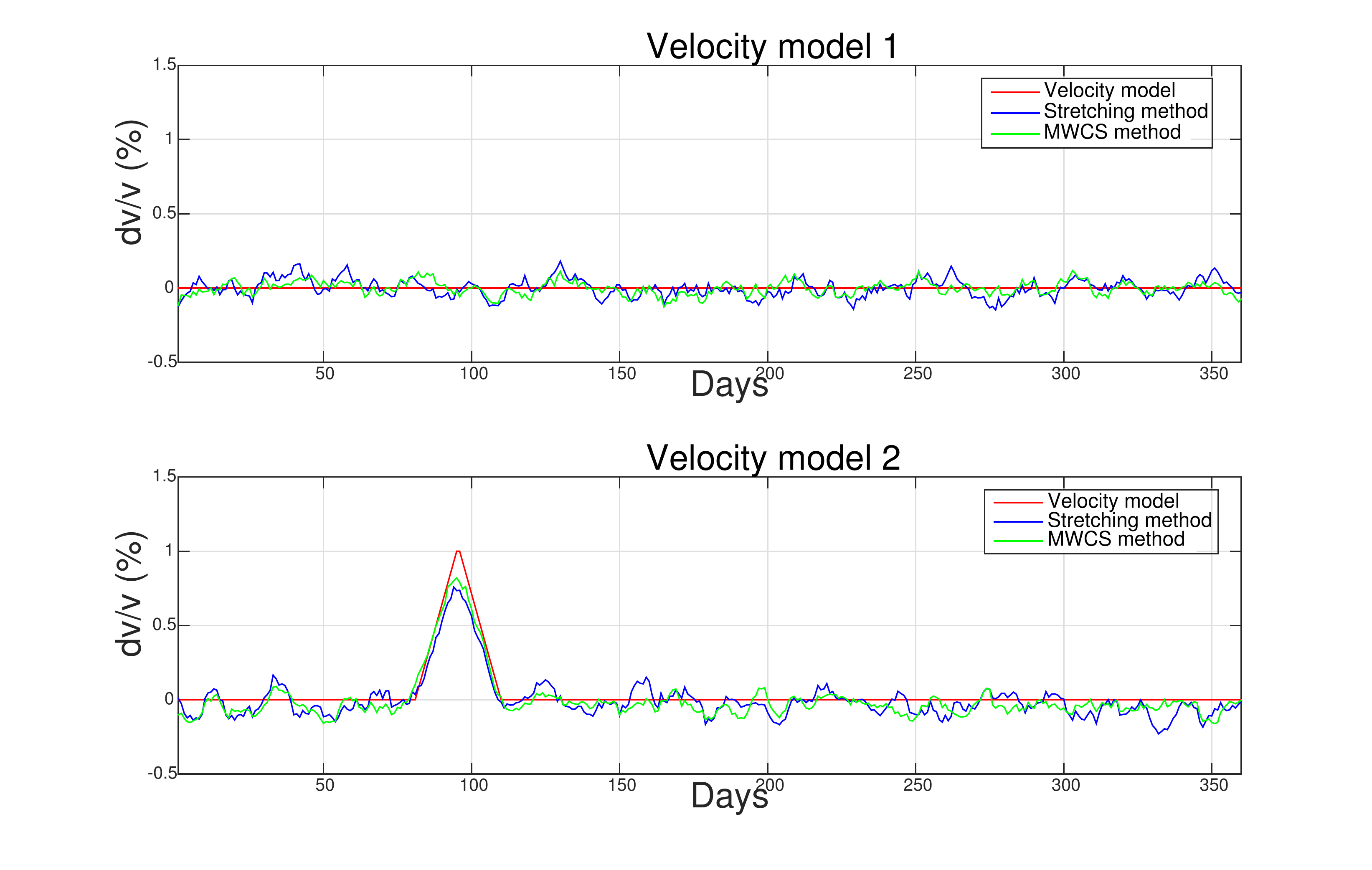}
\caption{Relative velocity change estimation using SM (blue) and MWCS (green) for the constant (top) and the variable (bottom) velocity models of Figure~\ref{fig:2}.}
\label{fig:num1}
\end{center}
\end{figure} 

The results obtained by both methods for the two velocity models are shown in Figure \ref{fig:num1}. We can see that the results are comparable 
and both methods can recover the relative velocity change up to a small error. We chose for the current cross-correlation a $N_{ccc}=7$-day average which minimizes the error in the estimation, as shown in Appendix \ref{ap:error} (see also Figure \ref{fig:App_error}).
 
\subsection{Seasonal variations in the noise sources and their influence to the relative velocity change measurements}
Let us write equation (\ref{eq:7}) in the frequency domain using equations (\ref{eq:2}) and (\ref{eq:4}),
\begin{equation}
\label{eq:11}
\begin{array}{l}
\widehat{\CC}^j(\omega,\mathbf{x}_1,\mathbf{x}_2)= \\[7pt]
\hspace{1cm}
\dsp \int d\by 
\; \overline{\hat{G}^j(\omega,\mathbf{x}_1,\mathbf{y})}\hat{G}^j(\omega,\mathbf{x}_2,\mathbf{y})
 \hat{\Gamma}^j(\omega, \mathbf{y}).
\end{array}
\end{equation}
Here $\omega \to \hat{\Gamma}^j (\omega,\mathbf{y})$ is the power spectral density of the noise sources at
location  $\mathbf{y}$ during day $j$ (see Appendix \ref{app:noisesources}).
As a complex function, the cross-correlation can be written as a product of an amplitude and a phase 
\begin{equation}
\label{eq:12}
\hat{\CC}_j(\omega,\mathbf{x}_1,\mathbf{x}_2)=A_j(\omega,\mathbf{x}_1,\mathbf{x}_2)e^{i\phi_j(\omega,\mathbf{x}_1,\mathbf{x}_2)}.
\end{equation}
We propose to use a normalization (spectral whitening) on the cross-correlation functions which consists in replacing the amplitude $A_j(\omega,\mathbf{x}_1,\mathbf{x}_2)$ by one in the frequency range where $A_j(\omega,\mathbf{x}_1,\mathbf{x}_2)$ 
is above a threshold.  Therefore we get, 
\begin{equation}
\label{eq:13}
\hat{\CC}_j(\omega,\mathbf{x}_1,\mathbf{x}_2)=e^{i\phi_j(\omega,\mathbf{x}_1,\mathbf{x}_2)}.
\end{equation}
After this spectral whitening we expect that the seasonal variations that affect only the amplitude spectra of the cross-correlation function 
will not have an impact on the measurement of $dv/v$. 

As shown in Appendix \ref{app:uniformornot}, when the seasonal variations of the noise sources are spatially uniform, then they affect only the amplitude spectra of the cross-correlations. Treating successfully the uniform case is important since we expect this hypothesis to be valid in most cases of interest where the receivers are close together geographically so that the seasonal variations are affecting in the same way, more or less, the ambient noise sources. 

However, if the seasonal variations affect also the phase spectra of $\CC$ then the spectral whitening will not ensure that the measurement 
of $dv/v$ will be free of apparent velocity changes due to seasonal variations of the noise sources. 
Our numerical model can simulate the daily perturbation of the power spectral density of the sources so as to be uniform or non-uniform with respect to the locations of the sources. 
The details of how this is implemented are in the appendix (see Appendix \ref{app:uniformornot}).

\subsubsection{Numerical simulations in a homogeneous medium}
We use here our numerical model with two different types of seasonal variations (uniform and non-uniform) and we study how these seasonal variations affect the estimations of the relative change in velocity when we use the stretching and the MWCS methods. 
We add first seasonal variations of a separable form as in equation (\ref{eq:15}). Then  (\ref{eq:11}) becomes
\begin{equation} \label{eq:star}
\hat{\CC}^j(\omega,\mathbf{x}_1,\mathbf{x}_2)= \hat{F}(\omega)\hat{f}^j(\omega)
 \int_{\cal C} d\sigma(\by) 
 \overline{\hat{G}^j(\omega,\mathbf{x}_1,\mathbf{y})}\hat{G}^j(\omega,\mathbf{x}_2,\mathbf{y})l(\mathbf{y}) ,
 \end{equation}
and we take first $l(\by)=1$. 

\begin{figure}
\begin{center}
\includegraphics[width=0.5 \textwidth]{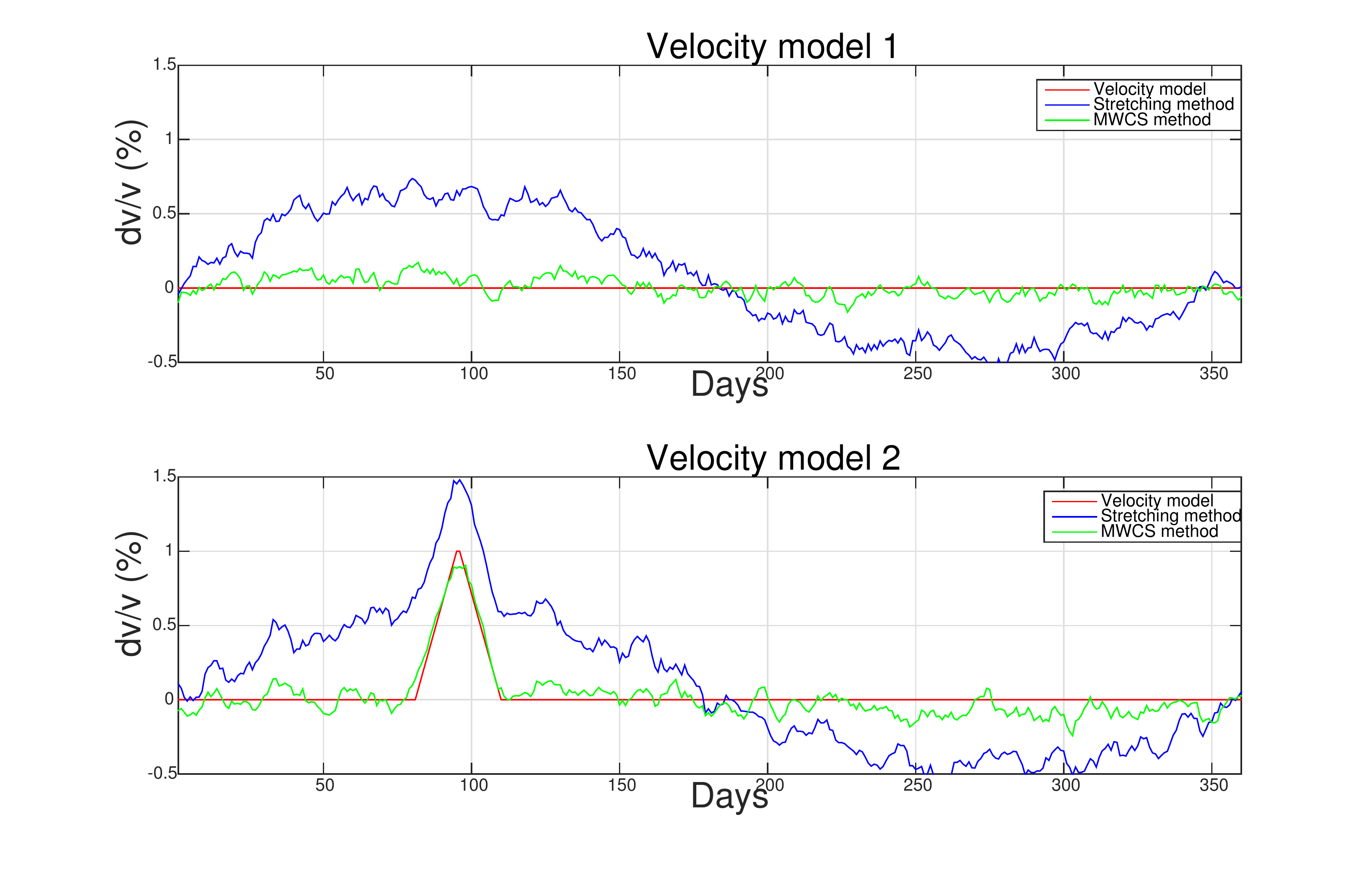}
\caption{Relative velocity change for the first (top) and the second (bottom) velocity model using SM (blue) and MWCS (green) for the velocity models of 
Figure \ref{fig:2}. Only the stretching method is affected by the seasonal variations since those are uniform with respect to the locations of the noise sources.}
\label{fig:num2}
\end{center}
\end{figure}

In this case only the amplitude of the cross-correlation is affected by the seasonal variations of the noise sources and therefore we expect only the stretching method to be affected. 
Indeed,  as we observe in Figure \ref{fig:num2} only the stretching method reflects the seasonal variations of the noise sources into seasonal variations on the measurement of $dv/v$. 
MWCS operates in the frequency domain and measures the phase difference between the two waveforms. Therefore,  seasonal variations in the amplitude spectra of the cross-correlation do not affect the MWCS estimation. 
 
\begin{figure}
\begin{center}
\includegraphics[width=0.5 \textwidth]{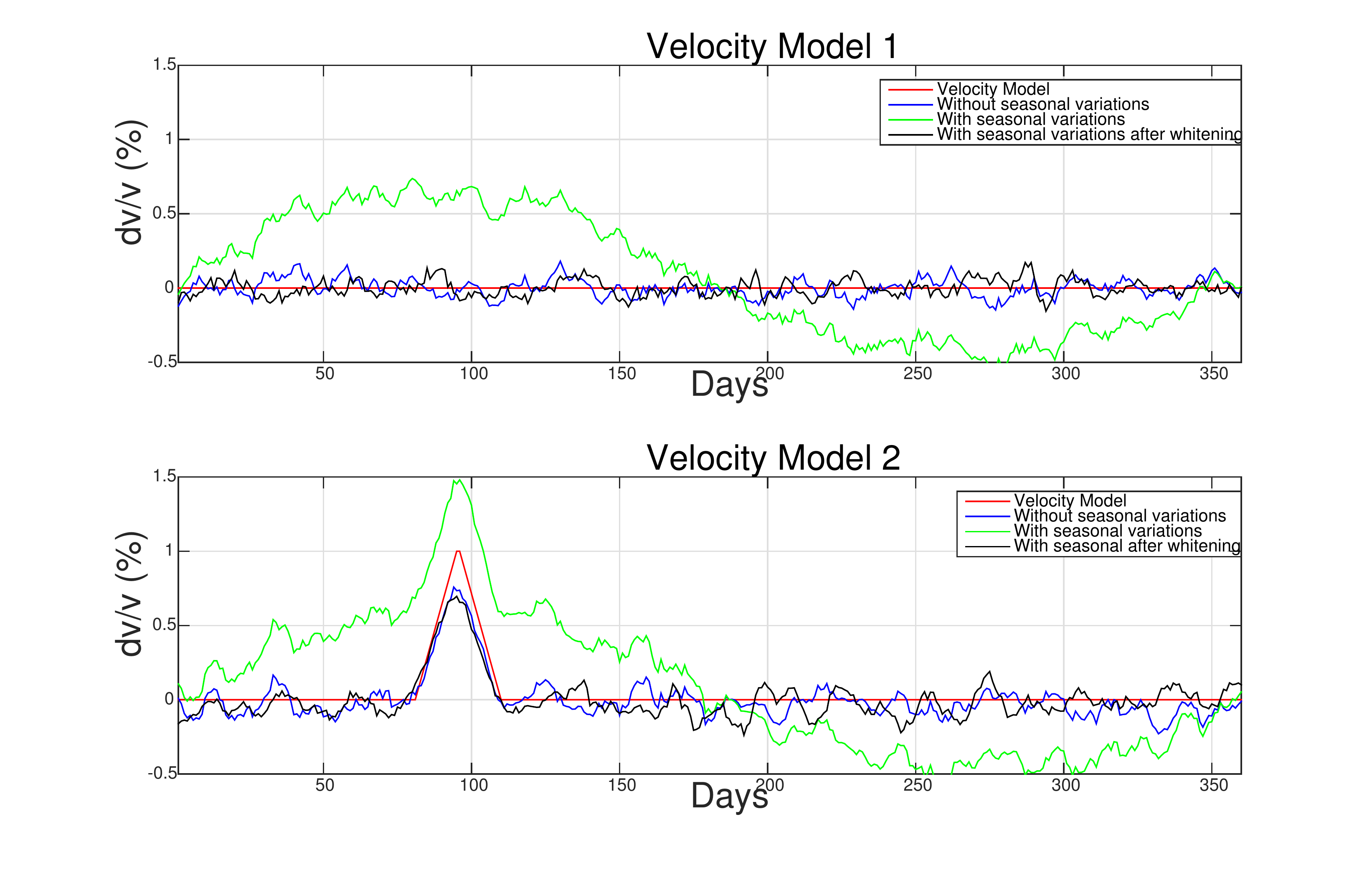}
\caption{Comparison between the estimation obtained for the model without seasonal variations in blue (equation (\ref{eq:6})), the model with uniform seasonal variations in green (equation (\ref{eq:15})) and the effect of spectral whitening to the estimation in black for both velocity models. All estimations here are produced using the stretching method.}
\label{fig:num3}
\end{center}
\end{figure}

By using spectral whitening we correct for the seasonal variations in the amplitude of the cross-correlation function and as a result we expect to no longer observe seasonal variations in the measurements of $dv/v$ when we use the stretching method. This is illustrated with our numerical results in Figure \ref{fig:num3}.

We do not expect to get the same result when the seasonal variations are of non-separable form as in equation (\ref{eq:16}). In this case, 
(\ref{eq:11}) becomes (for $l(\by)=1$)
$$
\hat{\CC}^j(\omega,\mathbf{x}_1,\mathbf{x}_2)= \hat{F}(\omega)
 \int_{\cal C} d\sigma(\by) 
 \overline{\hat{G}^j(\omega,\mathbf{x}_1,\mathbf{y})}\hat{G}^j(\omega,\mathbf{x}_2,\mathbf{y}))
  (1-\delta \hat{g}(\omega; \theta(\mathbf{y})+2\pi j / N_d)\sin(2\pi j /N_d))^2
  ,
$$
where $\theta(\mathbf{y})$ is the angle of $\mathbf{y}$ on the circle ${\cal C}$, $\delta=0.4$ and $\hat{g}$ is defined in the Appendix (see \eqref{eq:16b},\eqref{eq:17})

\begin{figure}
\begin{center}
\includegraphics[width=0.5 \textwidth]{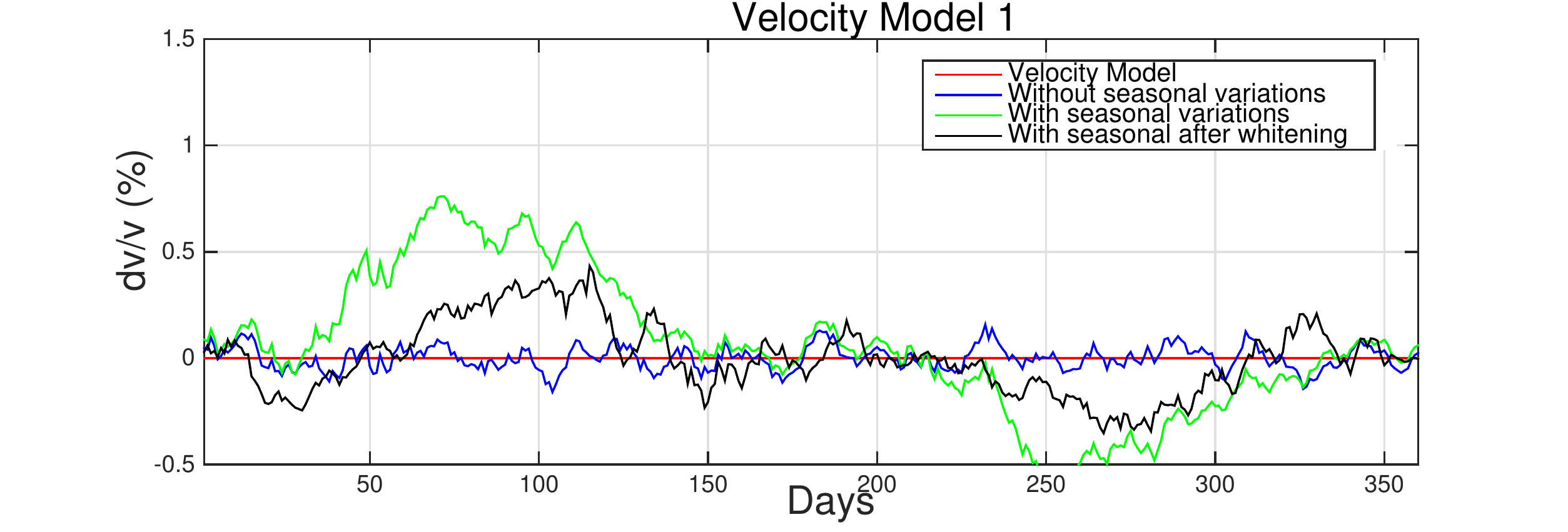}
\caption{The estimation produced by the stretching method for  the numerical model without seasonal variations in blue (equation (\ref{eq:6})), the model of uniform seasonal variations in green (equation (\ref{eq:16})) and the effect of spectral whitening to the estimation in black.}
\label{fig:num4}
\end{center}
\end{figure}

Indeed, we as we observe in Figure \ref{fig:num4}, spectral whitening cannot remove the seasonal variations any longer since those variations affect both the amplitude and phase spectra of the cross-correlation.

\subsubsection{Simulations in a scattering medium}
\begin{figure}
\begin{center}
\includegraphics[width=0.5 \textwidth]{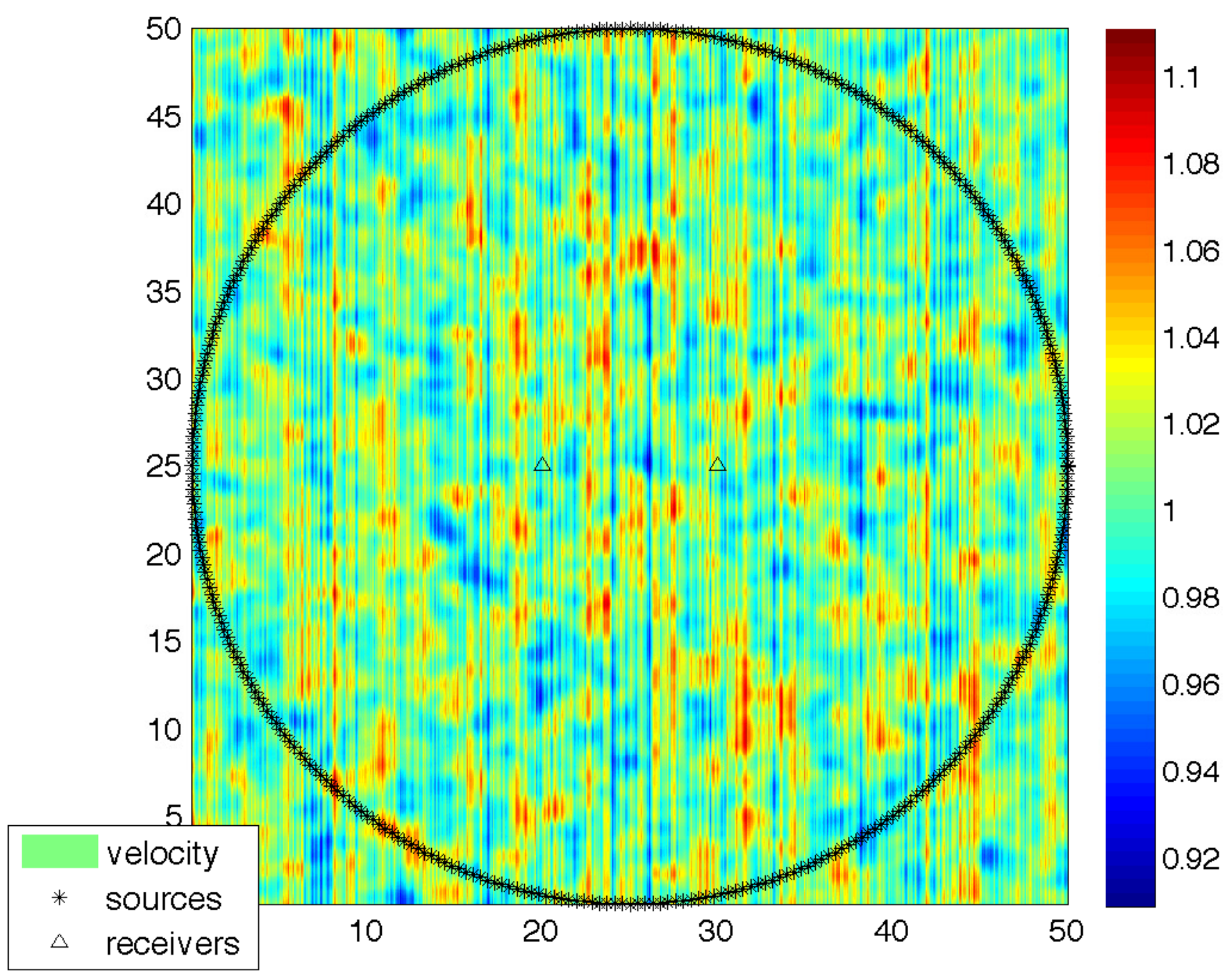}
\caption{Highly scattering medium. The positions of the sources/receivers are the same as in the homogeneous case (see Figure \ref{fig:1}). }
\label{fig:mediumNEW}
\end{center}
\end{figure}

The results presented in the previous section are for a homogeneous medium and are extracted using the direct waves in the cross-correlations. More precisely we used the time window $[10.5,20.5]$s (travel-time between the sensors$=10s$). To illustrate the generality of our approach we consider here the case of a scattering medium. The Green's function is computed now by solving the wave equation in a square domain of 
$50$Km$\times50$Km (see Figure \ref{fig:mediumNEW}) filled with a scattering medium with an average velocity of $1$Km/s and $10\%$ fluctuations. 
The medium fluctuations  here may produce less scattering than the circular inclusions with a contrast of 50\% considered in \cite{colombi} but our fluctuations model seems quite realistic in the geophysical context.
The wave equation is solved with the software Montjoie \href{http://montjoie.gforge.inria.fr/}{(http://montjoie.gforge.inria.fr/)} 
using seventh order finite elements for the discretization in space and fourth order finite differences in time. The computational domain is surrounded by a perfectly 
matched absorbing layer model (PML). 

\begin{figure}
\begin{center}
\includegraphics[width=0.45 \textwidth]{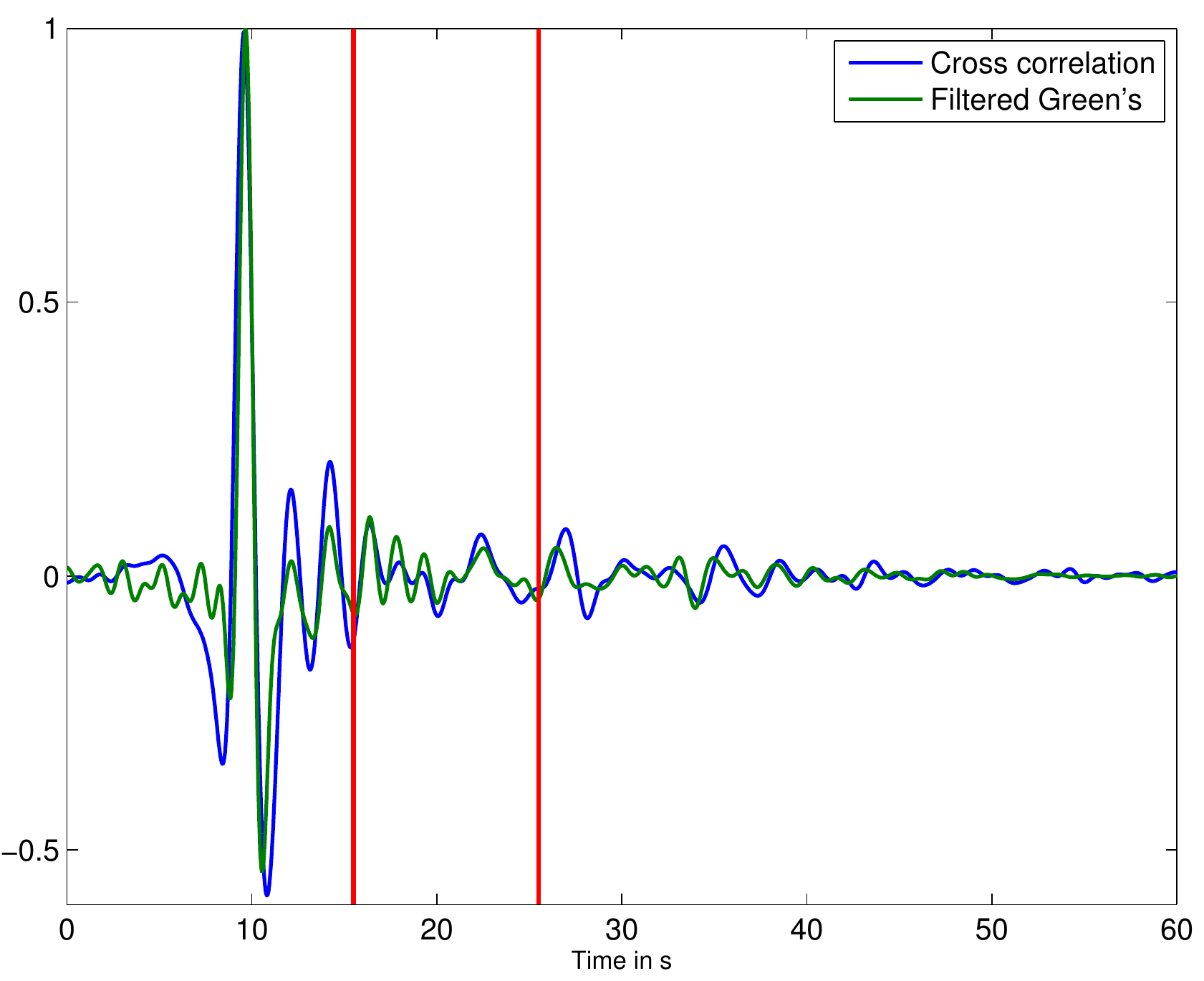}
\includegraphics[width=0.45 \textwidth]{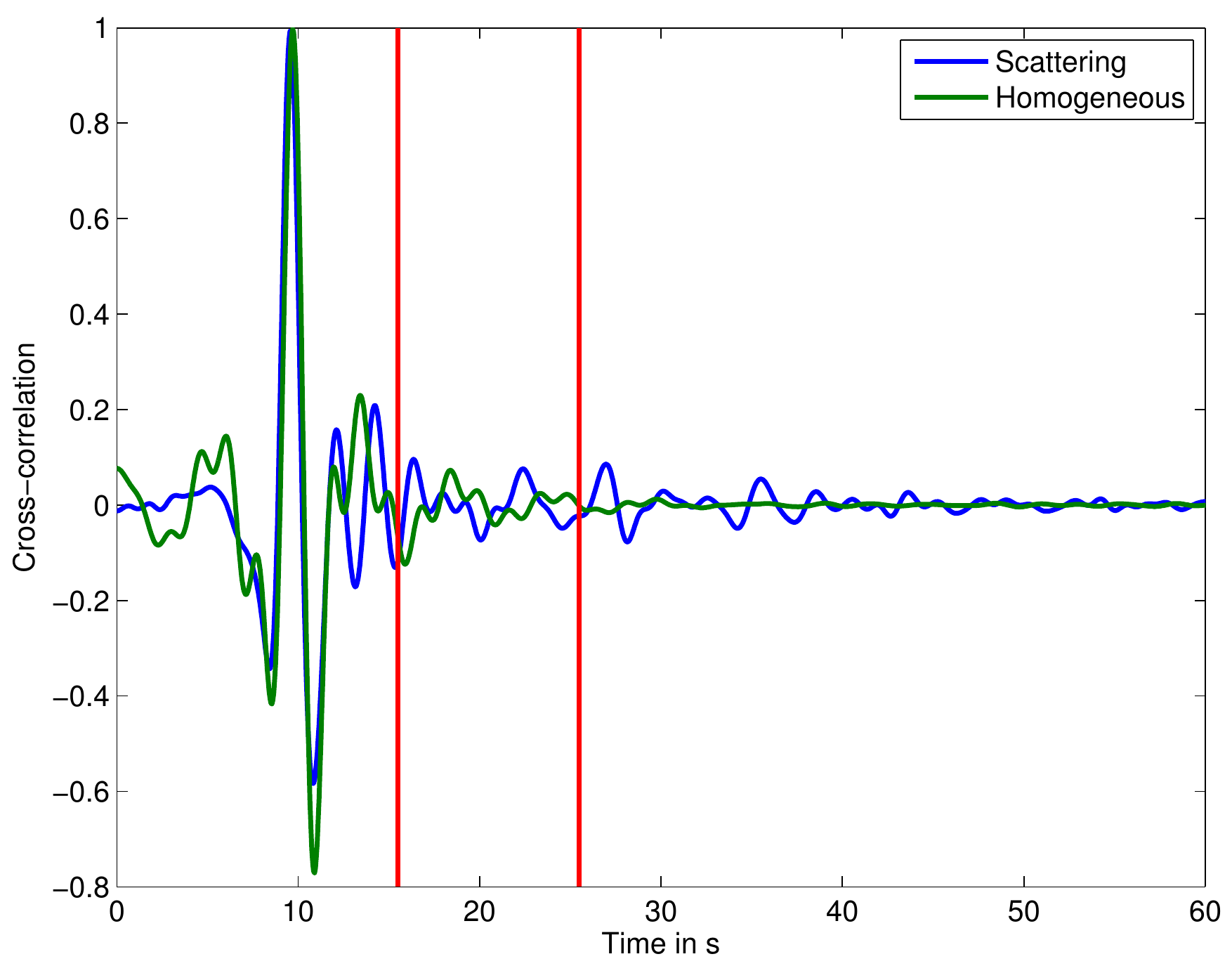}
\caption{Left: The reference $\CC$ in the scattering medium compared with the Green's function between the two receivers filtered by the power spectral density of the noise sources. Amplitudes are normalized.  Right: The reference $\CC$ in the homogeneous and the scattering medium. In both plots, the two red vertical lines indicate the window $[15.5-25.5]$s. }
\label{fig:CCscat}
\end{center}
\end{figure}

\begin{figure}
\begin{center}
\includegraphics[width=0.49 \textwidth]{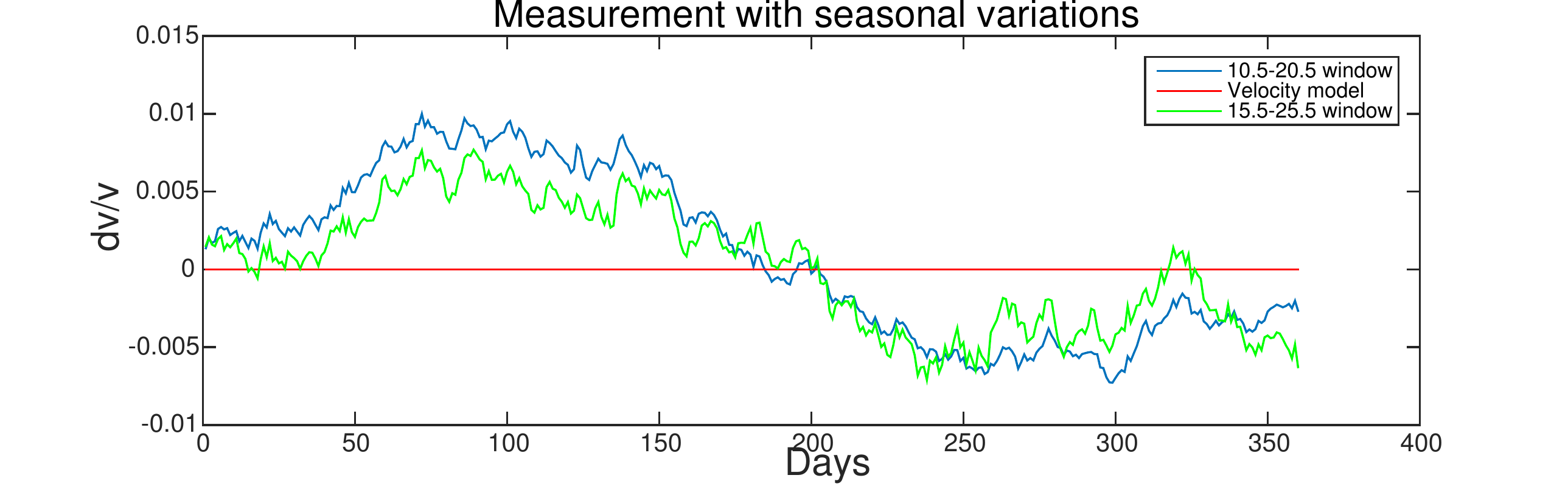}
\includegraphics[width=0.49 \textwidth]{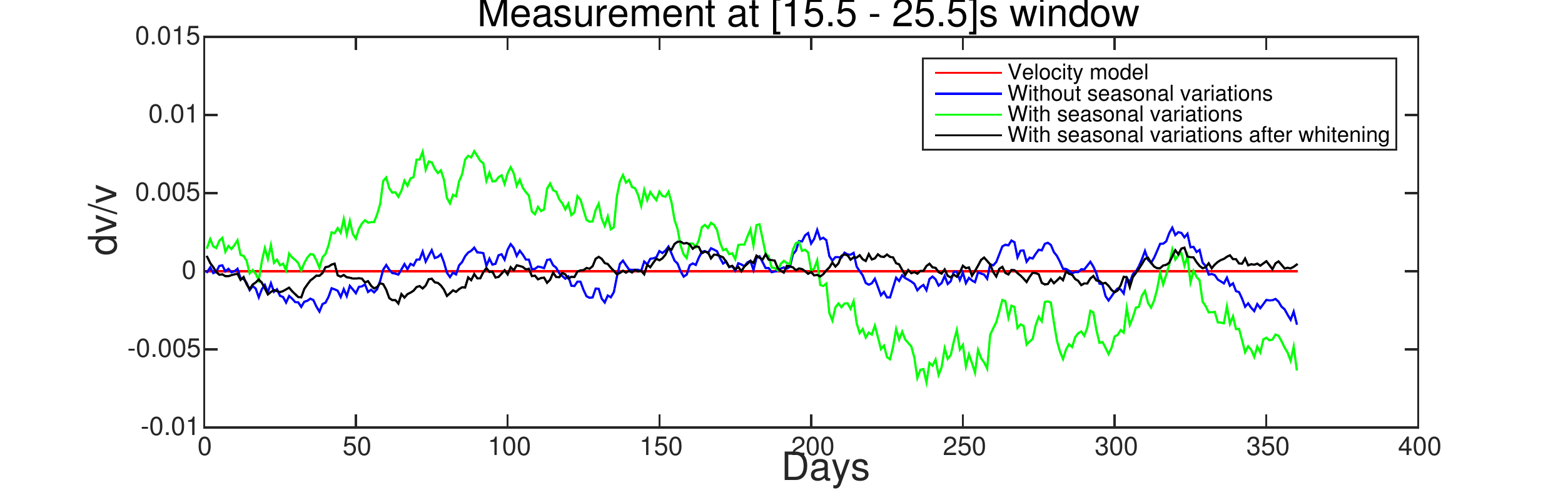}
\caption{Scattering medium. Left: SM estimation of $dv/v$ in the present of seasonal variations of a separable form using two different time windows. Right: The seasonal variations are removed using spectral whitening (here the measurements are performed with the $[15.5-25.5]$s window).}
\label{fig:Coda}
\end{center}
\end{figure}
In Figure \ref{fig:CCscat}-left we compare the reference $\CC$ function with the Green's function between the two receivers obtained by emitting a pulse from one receiver and recording it at the other. We observe a very good agreement between the two signals up until $\approx 42$s. In Figure \ref{fig:CCscat}-right we compare the reference $\CC$ function in the scattering medium with the one in the homogeneous medium. The oscillations before and after the main peak of the pulse in the homogeneous medium are due to the limited bandwidth of the noise sources. Note that the two signals differ significantly after 12.5s.

We consider now seasonal variations of separable form as in 
\eqref{eq:star} with $l(\by)=1$ and estimate $dv/v$ with the stretching method using two different time windows: first the same window as before $[10.5-20.5]$s, and second, the window $[15.5-25.5]$s. 
As we can see at Figure \ref{fig:Coda}-left the apparent false variations in $dv/v$ are reduced 
by using the coda part of the $\CC$ but they still persist. The proposed spectral whitening of $\CC$ efficiently removes the fluctuations as illustrated  in Figure \ref{fig:Coda}-right. 
Let us emphasize that spectral whitening will be efficient for any spatio-temporal variation of the noise sources of separable form 
since such variations affect only the amplitude of $\CC$ and this regardless of the underlying medium (homogeneous or scattering). 

\subsubsection{Simulations for anisotropic noise distributions}
\begin{figure}
\hspace*{4cm} Homogeneous \hspace*{5cm} Scattering \\
\raisebox{1.8cm}{MWCS} \includegraphics[width=0.45 \textwidth]{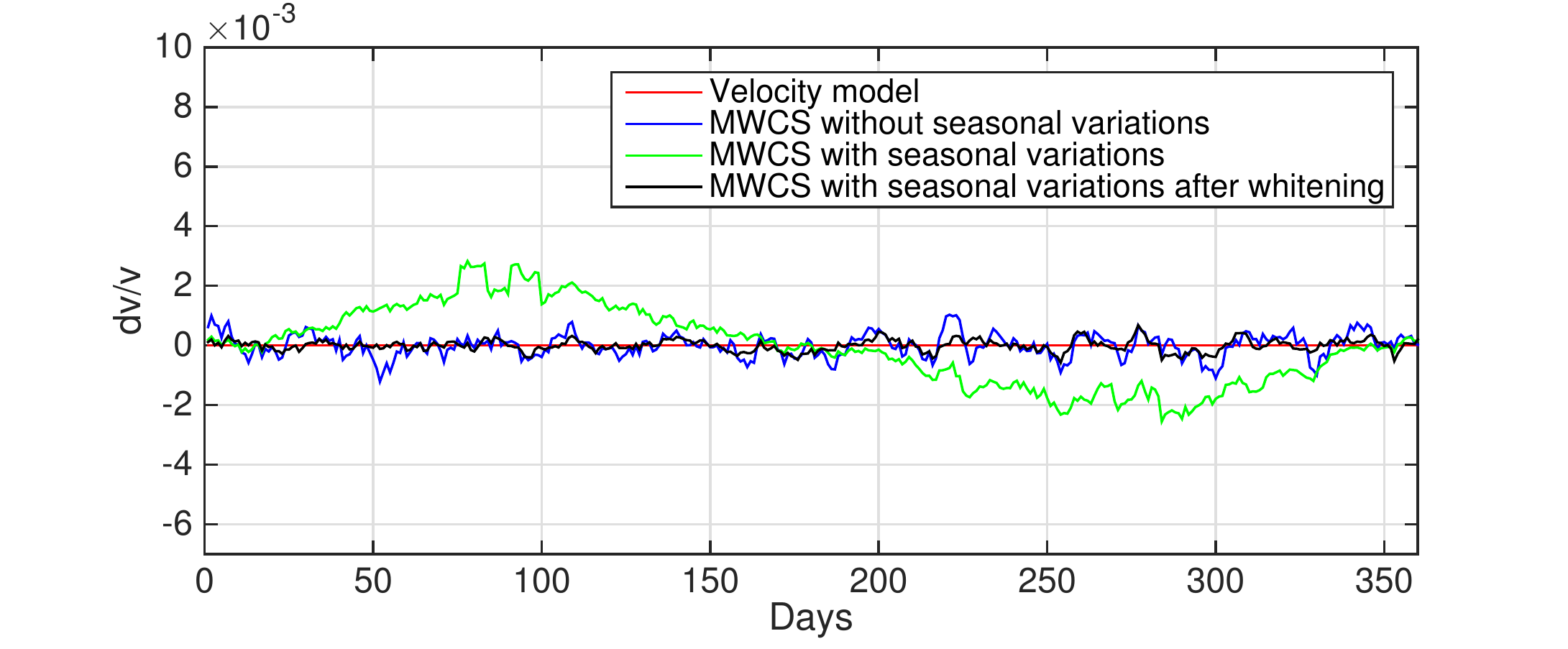} \hspace{-0.7cm}
\includegraphics[width=0.45 \textwidth]{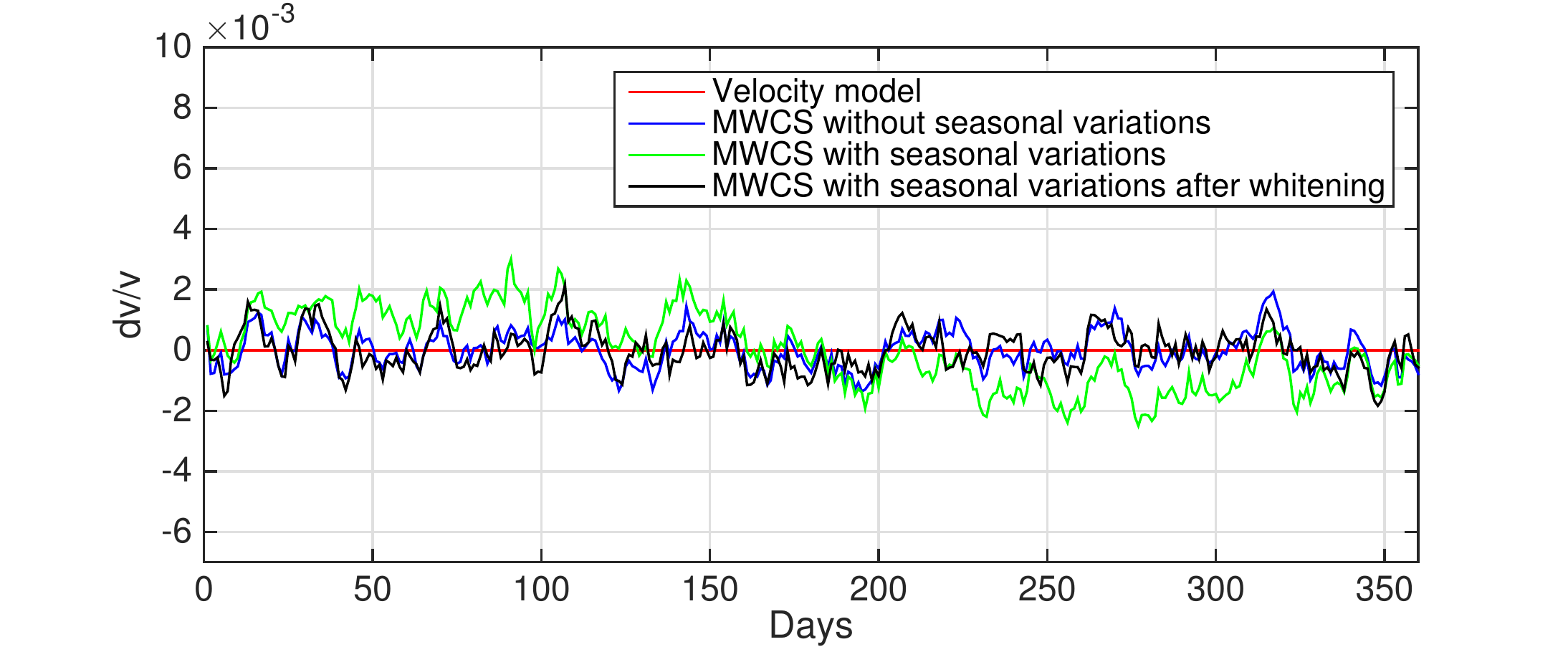}\\
\hspace*{0.6cm} \raisebox{1.8cm}{SM}\includegraphics[width=0.45 \textwidth]{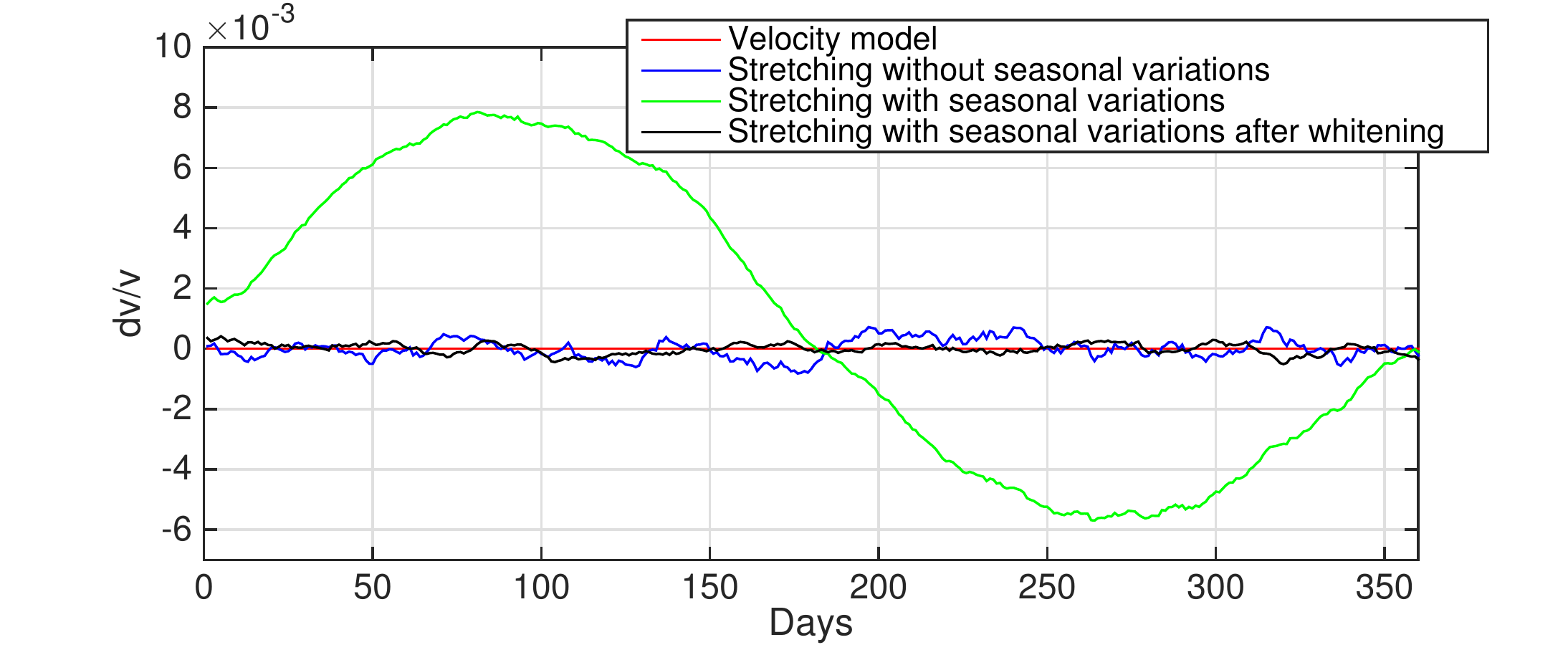}\hspace{-0.7cm}
\includegraphics[width=0.45 \textwidth]{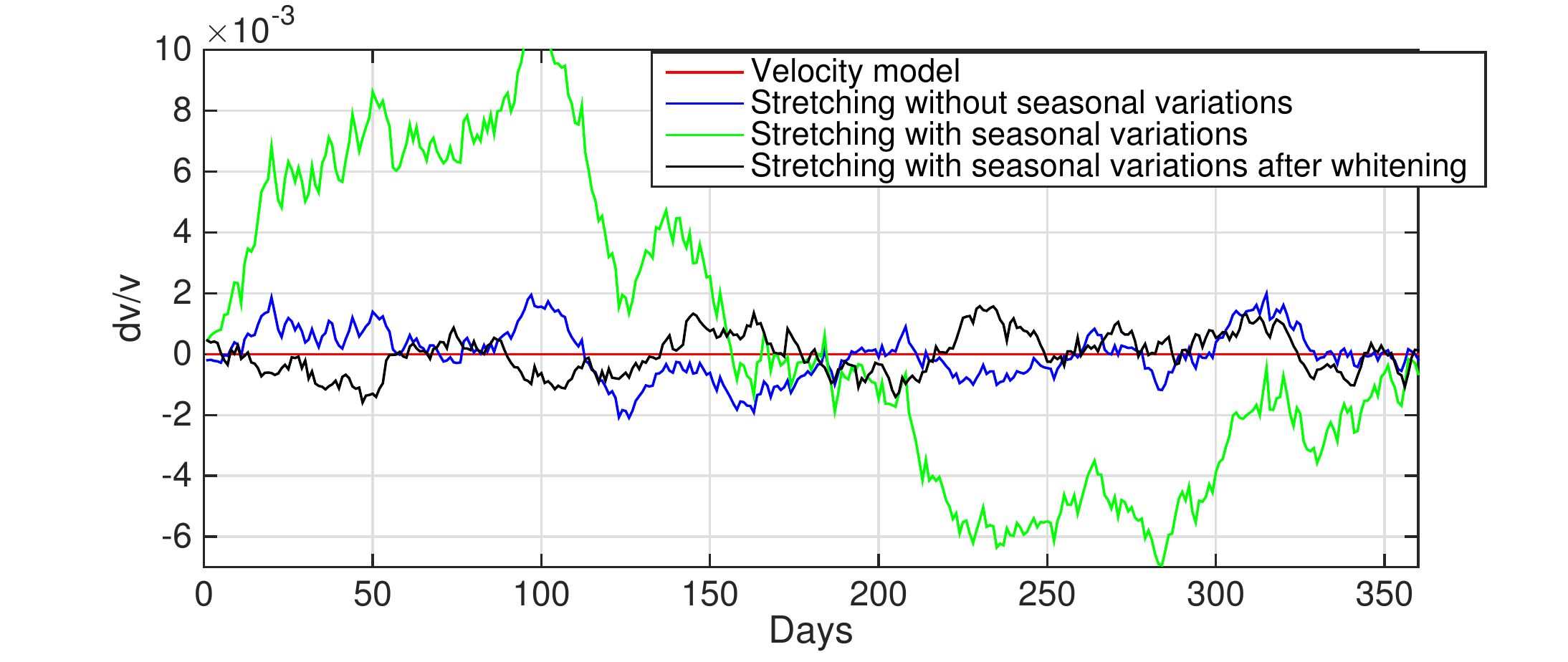}
\caption{Removing the seasonal variations using spectral whitening. The noise sources have anisotropic spatio-temporal fluctuations as described by \eqref{eq:15b}-\eqref{sjanis}.
For the SM method, the measurements in the homogeneous medium are performed using the $[10.5-20.5]$s window while in the scattering medium the $[15.5-25.5]$s window is used.}
\label{fig:anis}
\end{figure}
To further illustrate the robustness of the proposed filtering 
we add now anisotropy to the noise sources. Following \cite{colombi} we consider a rather extreme case of anisotropy using \eqref{eq:15b} which amounts to cross-correlations as in \eqref{eq:star} with azimuthal intensity distributions of the form, 
$$l(\by)= (1-0.6\cos{(2\theta(\by))})^2,$$
with $\theta(\mathbf{y})$ the source azimuth, {\em i.e.}, the angle of $\mathbf{y}$ on the circle ${\cal C}$.
The results obtained with MWCS and SM in homogeneous and scattering media before and after spectral whitening are shown in Figure \ref{fig:anis}. 
As expected the MWCS estimation is less affected by the spatio-temporal variations of the noise sources since to the leading order the phase of the cross-correlation remains unchanged \cite{weaver09}. The amplitude of the cross-correlation however is affected and this leads to erroneous estimates with SM. The results of both methods are greatly improved with spectral whitening. 
In the scattering medium the anisotropy effect of the noise sources is alleviated through the multiple scattering of the waves with the medium inhomogeneities.  This corrects for the anisotropy effect on the phase of the cross-correlation but not on the amplitude. Therefore SM estimation remains bad while the MWCS estimation is better in the scattering medium. Again the results of both methods are improved with spectral whitening.

\subsection{Seasonal variations examined in the island of Milos}
\begin{figure}
\begin{center}
\includegraphics[width=0.5 \textwidth]{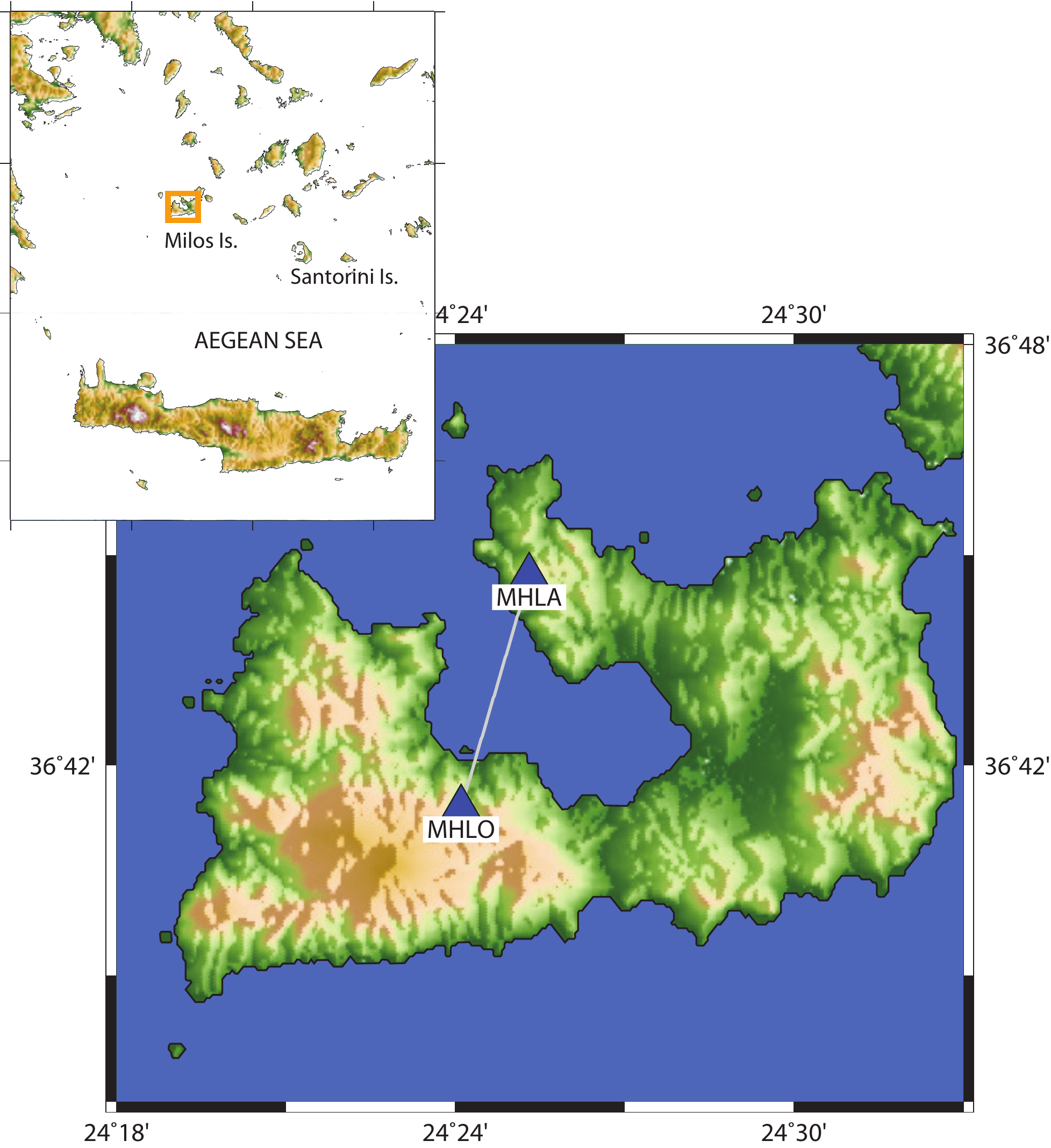}
\caption{The volcanic island of Milos and the locations of the two NOAIG broadband seismic stations used in this study. The inset at the left hand side of the map shows the location of Milos island (orange rectangle) within the Aegean sea.}
\label{fig:rays_milos}
\end{center}
\end{figure}

\begin{figure}
\begin{center}
\includegraphics[height=6cm]{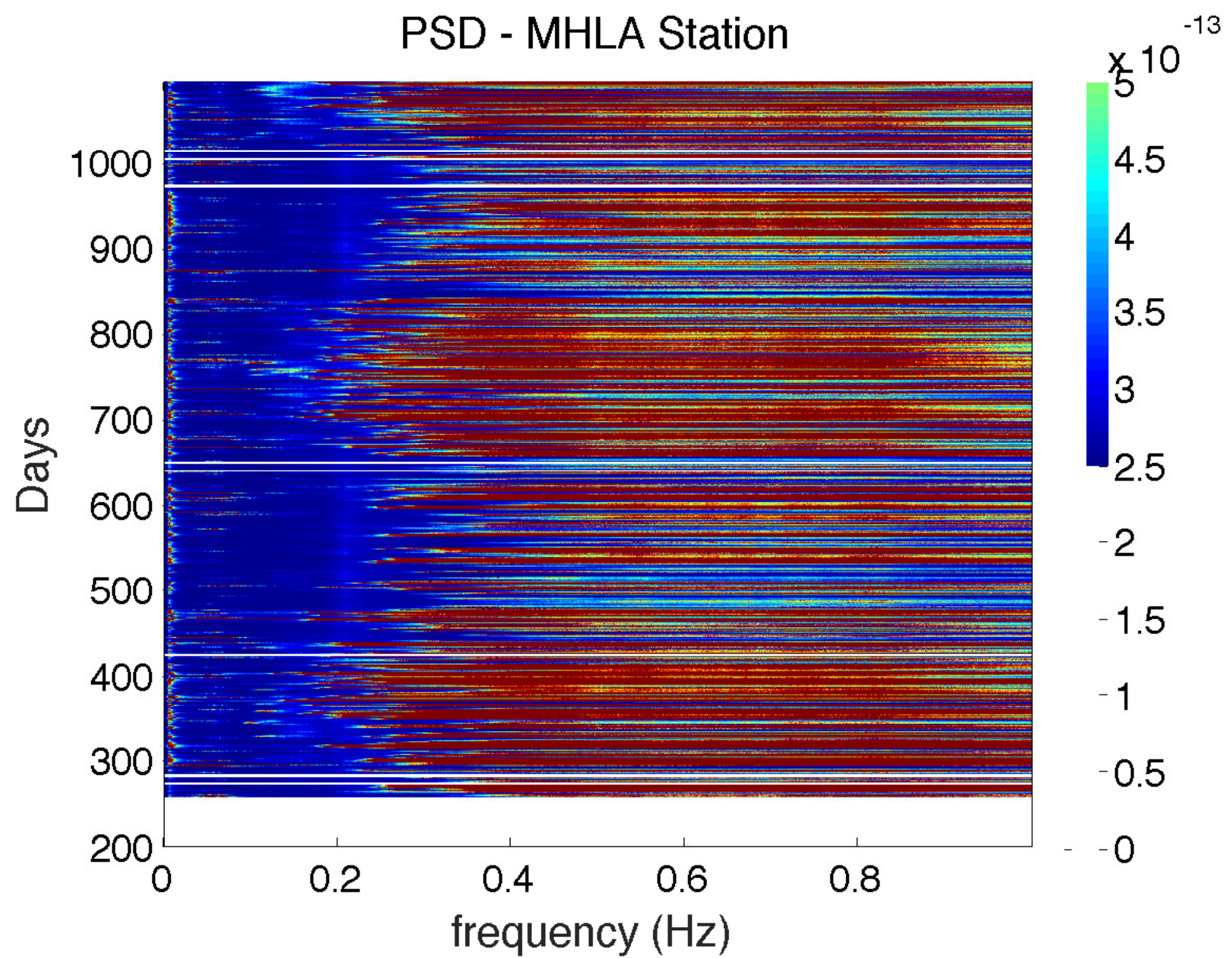}
\raisebox{0.4cm}{\includegraphics[height=5.5cm]{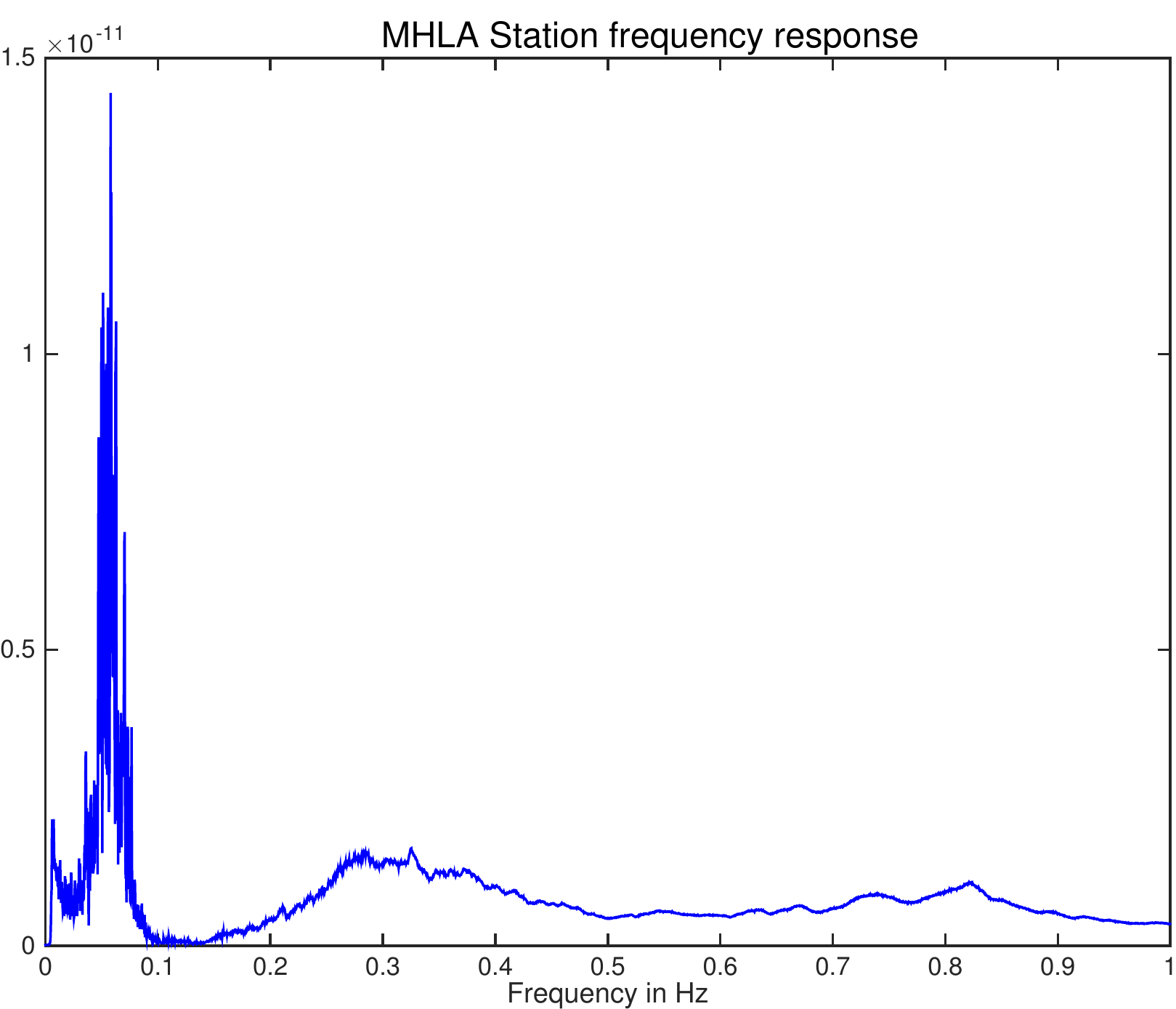}}
\caption{Left: The Power Spectrum Density of  the station MHLA at Milos. Right: The frequency response of the MHLA station calculated by averaging the daily frequency response of all available days.}
\label{fig:2D_Milos}
\end{center}
\end{figure}

Using the developed methodology we investigate here relative velocity changes in the quiet volcanic island of Milos, in Greece. In the area two broadband seismic stations (codes: MHLO and MHLA) operate in real time, monitoring seismicity in the Aegean volcanic arc for the National Observatory of Athens, Institute of Geodynamics (NOAIG) (Figure \ref{fig:rays_milos}). The two stations are part of the Hellenic National Seismic Network (network code: HL) and they are deployed 6km apart and above the Milos island geothermal reservoir. 
We gather seismic noise recordings for the last days of 2011 and the entire 2012 and 2013  (827 days in total). During the examined period there was no significant local earthquake activity 
in the area. In Figure \ref{fig:2D_Milos}-left we observe the seasonal variations on the Power Spectrum Density (PSD) of the station MHLA and we want to investigate if the stretching method is affected by those variations. These seasonal variations have been attributed to local sea–weather conditions within a range of a few hundred kilometers from the stations \cite{EvangelidisM12}.

The data are filtered from $0.1-1.0$Hz a bandwidth for which we have microseismic activity as suggested by Figure \ref{fig:2D_Milos}-right. This frequency bandwidth will be used for Santorini in the next section since the power spectral density of the recorded signals is more or less the same. 
 
\begin{figure}
\begin{center}
\includegraphics[width=0.5 \textwidth]{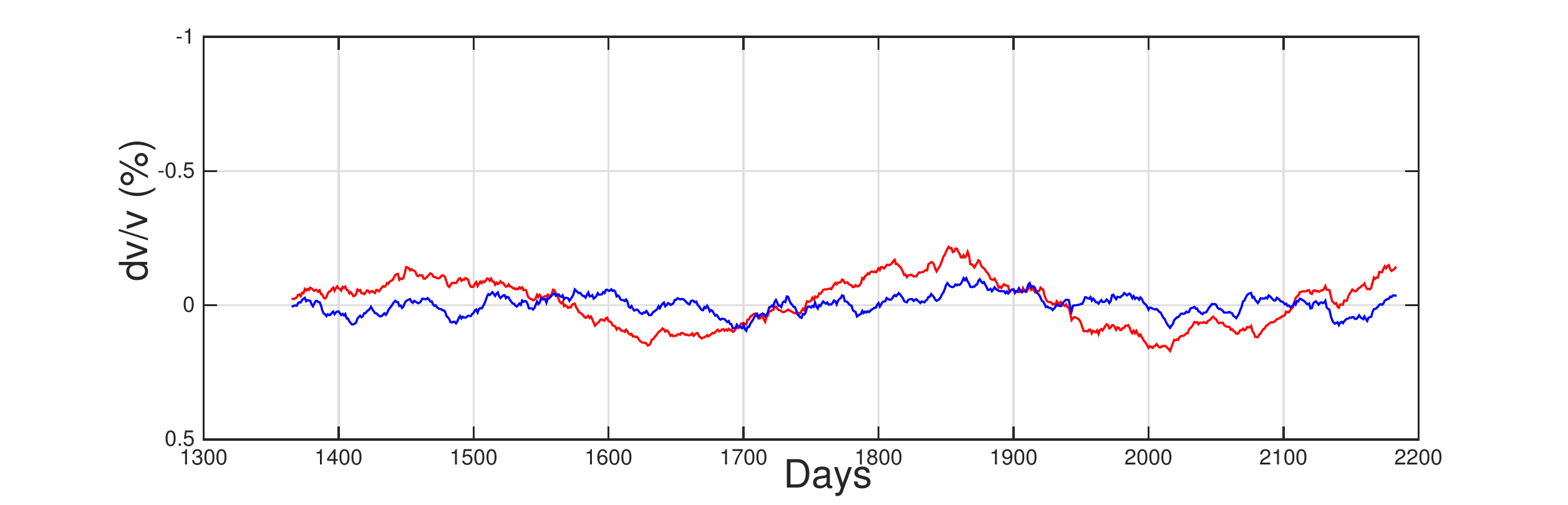}
\caption{The estimation between the pair MHLO-MHLA located on the island of Milos when we use spectral whitening (blue) and when we do not use it(red). Here $N_{ccc}=21 $ Days. }
\label{fig:real_s}
\end{center}
\end{figure}

As we see in Figure \ref{fig:real_s}, the proposed normalization (spectral whitening) has the desirable effect on seasonal variations just as the numerical simulations suggest. 
Considering the apparent velocity fluctuations induced by seasonal variations of the noise sources, as measurement noise, we obtain a decrease in the noise level of the order of 3 after using the proposed normalization.
Using the stretching method with spectral whitening, we observe residual fluctuations in the estimated velocity of the order of $\pm 0.1\%$.

\section{Investigation of the Santorini island seismic unrest 2011-2012} \label{santorini}
\begin{figure}
\begin{center}
\includegraphics[width=0.5 \textwidth]{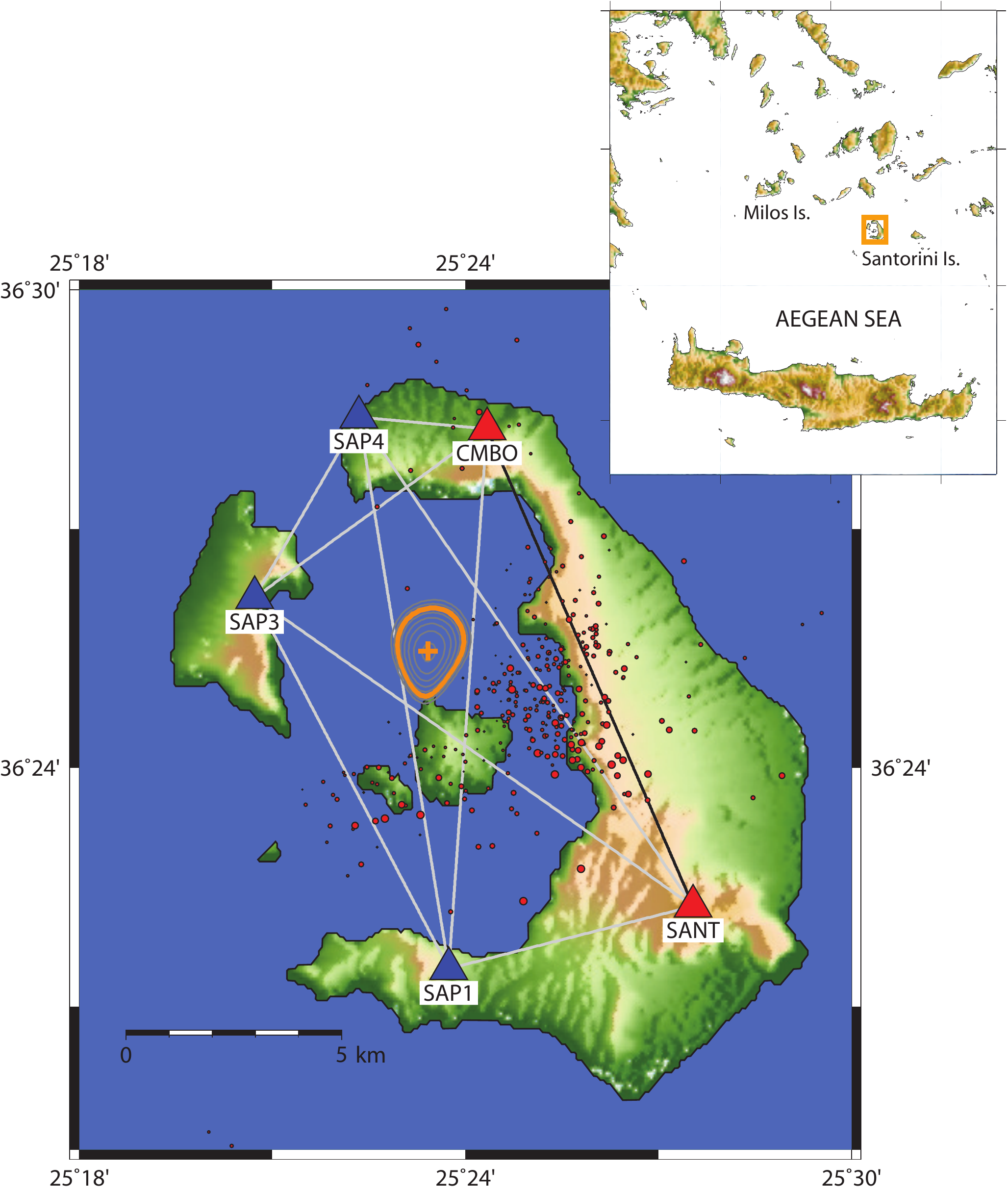}
\caption{Network of seismic stations in Santorini and the inter-station paths.  Stations that were in operation prior the unrest are marked in red. Stations that became operational during or after the unrest are marked in blue. Red circles indicate the relocated seismicity according to \cite{Kwnst2013} with their size being proportional to the event local magnitude ($M_L$) as measured by NOAIG. The orange cross marks the geographic location of the modeled volumetric growth at 4~km depth \cite{Newman} with their $95\%$ confidence level (concentric circle). The inset at the right hand side of the map shows the location of Santorini island (orange rectangle) within the Aegean sea.}
\label{fig:3}
\end{center}
\end{figure}

During the time period January 2011 to March 2012, high microseismic activity was observed in the caldera of the Santorini island (Figure \ref{fig:4}). This also coincided with a 10cm uplift measured by GPS stations deployed in the area, monitoring continuously crustal deformation \cite{Newman}. During the unrest period, several portable seismic stations were deployed in the area by research institutions and universities. However, due to the urgency of the ongoing unrest, the portable stations were deployed mainly to monitor seismicity in near real time and thus, their data quality and/or availability was not suitable for ambient noise monitoring. Prior to the unrest, only two digital broadband seismic stations were in operation (Figure \ref{fig:3}). These two were found useful to use for investigating variations in $dv/v$ using the stretching method. Their inter-station path crosses the edge of the uplifted area within the caldera which is also the source region of the majority of the observed seismic clusters \cite{Kwnst2013}. \\ %

The unrest was studied in \cite{Lagios2013} and \cite{timespace2014}  using  GPS data and the results suggest elevation at the volcano mainly at periods with high seismicity. More specifically the seismic activity was high from January 2011 until August 2011 and then it is high again from October 2011 to February 2012.  Those two periods of high seismicity are the same periods during which  GPS data suggest that there is an elevation of the caldera.

\subsection{Data Treatment}
For each pair of stations we follow the next steps. First we separate the 24-hours long segment of each station into eight 3-hours segments. If a 3-hours long segment has more than $10\%$ of gaps then it is rejected and will not be used in the calculations of the cross-corellation (\CC). 
Otherwise, we filter the data in the band $[0.1-1.0]$Hz. Then we apply one-bit normalization and we cross-correlate with the corresponding segment from the paired station. For each day we expect at most eight Cross-Correlation functions. If a 3-hour segment is rejected then we miss one cross-correlation and only if for one day we miss three or less cross-correlation functions we proceed 
and average the 3-hours segments to get the daily cross-correlation function. A final step that helps us to deal, under some conditions,  with seasonal variations in the power spectral density of the noise sources, is to apply spectral whitening on the cross-correlations inside the bandwidth of interest, i.e., $[0.1-1.0]$Hz. \\
\begin{figure}
\begin{center}
\includegraphics[width=0.4 \textwidth]{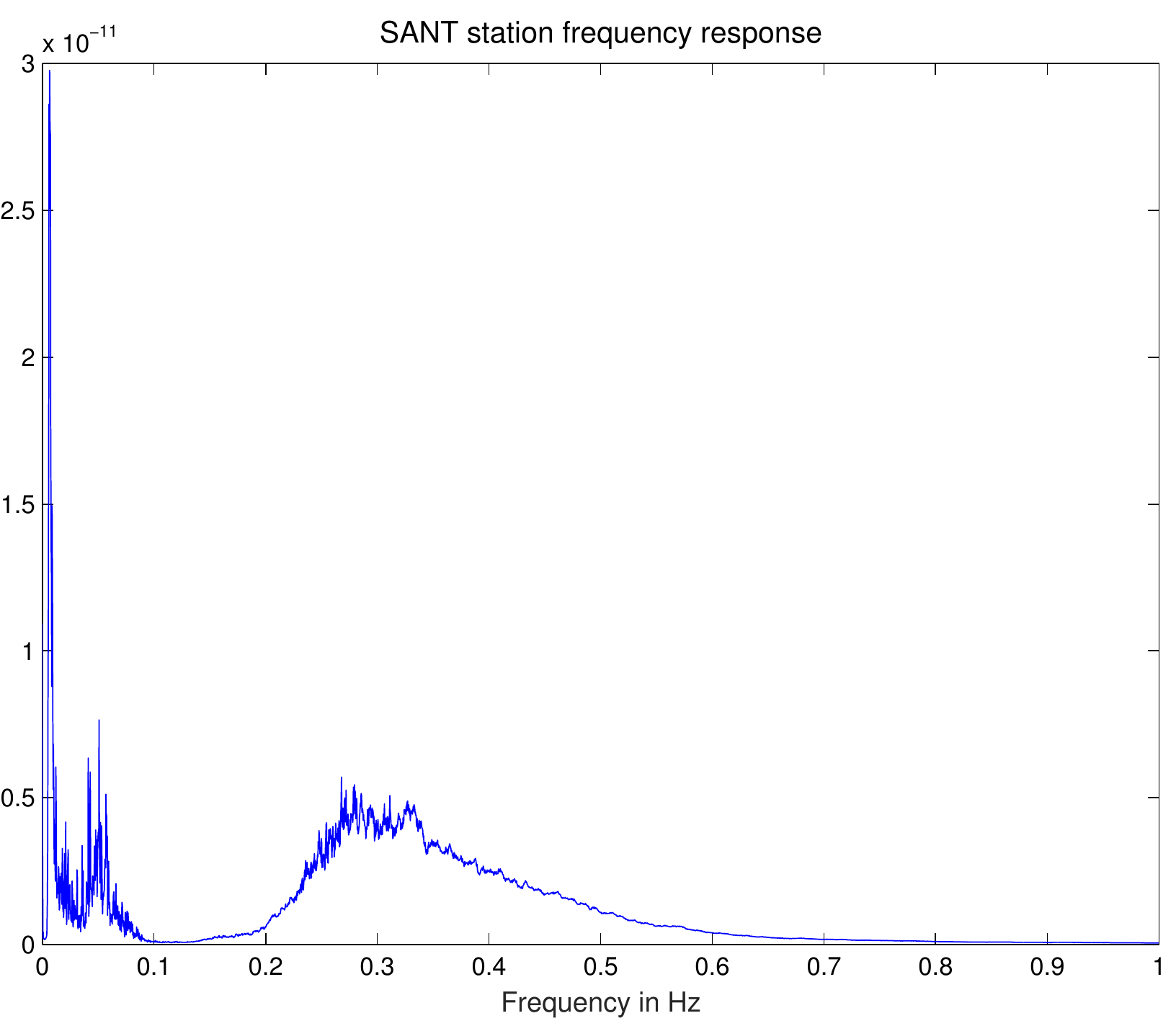}
\caption{The frequency response of the SANT station calculated by averaging the daily frequency response of all available days.}
\label{fig:1D_SANT}
\end{center}
\end{figure}

For the reference cross-correlation function we use the mean of all available daily cross-correlation functions. The current cross-correlation function on the other hand is the mean of $N_{ccc}=21$ days around the day where we want to make the measurement.

\subsection{Results}

Our implementation of the stretching method is configured to make two measurements of $dv/v$ using the positive and the negative time axis in a time window that is focused on the coda part.
($[15,35]$s and $[-35,-15]$ in our case). The final result is the average of the two measurements as long as the correlation coefficient is higher than 0.7 otherwise the result is rejected.\\

\begin{figure}
\begin{center}
\includegraphics[width=0.5 \textwidth]{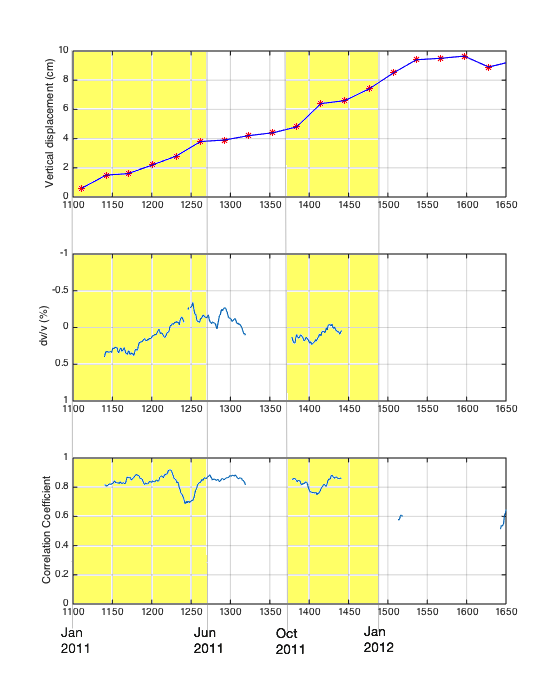}
\caption{Top: Accumulated elevation of the GPS station NOMI  in Santorini \cite{timespace2014}. Middle: The estimation of $dv/v$ using the stretching method. Bottom: The correlation coefficient of the stretching method.}
\label{fig:4}
\end{center}
\end{figure}

The drop of the $dv/v$ is maximal in May 2011, associated with a considerable drop of the CC coefficient (Fig. \ref{fig:4}). This implies a change in the scattering medium at least for these days.

Unfortunately we do not have data that cover the entire period of the unrest but as we can see in Figure (\ref{fig:4}) we can compare the available data with GPS data (from the GPS station NOMI, located roughly in the middle of the inter-station path between SANT and CMBO). 

\begin{figure}
\begin{center}
\includegraphics[width=0.5 \textwidth]{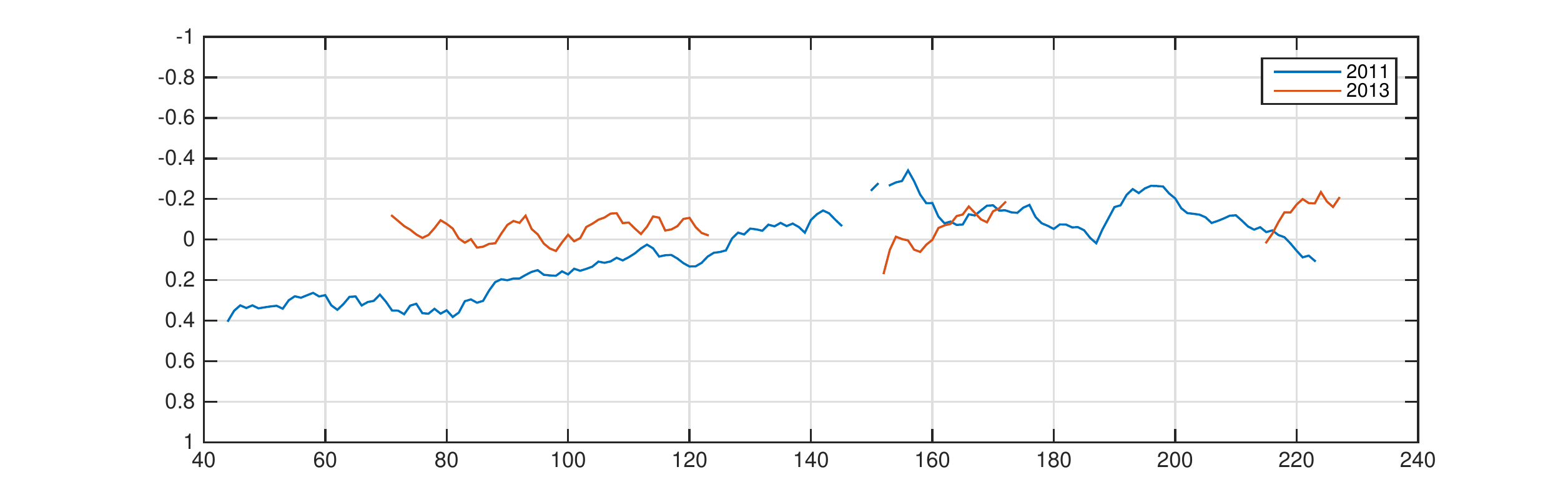}
\caption{Results using Stretching method in two different years, for Julian dates between 40 and 240.}
\label{fig:4_J}
\end{center}
\end{figure}

The result shown in Figure \ref{fig:4} middle plot is quite close to the GPS measurements, at least during the periods that we have available data and for the periods with high seismic activity (high seismic activity corresponds to the yellow background). We can also see that the elevation increases mainly at the periods of high seismic activity according to the GPS data (top plot at Figure \ref{fig:4}).\\ 
Based on the data for Milos (Figure \ref{fig:real_s}) and for Santorini in 2013 (Figure \ref{fig:4_J} ,red),
the estimated velocity has random fluctuations of the order of $\pm 0.1\%$, resulting from residual seasonal variations and errors in the estimation. 
Therefore, any change of more than $\pm 0.1  \%$ can be considered as significant, {\em i.e.}, resulting from physical changes in the velocity distribution.
This is what happens in Santorini in 2011 (Figure \ref{fig:4_J}, blue).

\section{Conclusions}
In this paper we considered the problem of seismic velocity change estimation based on passive noise recordings. Using simple 
but realistic numerical simulations as a tool, we study how the estimation produced by the stretching method is affected by seasonal 
spatio-temporal fluctuations of the amplitude spectra of the noise sources \cite{clayton,labasin}. Moreover, 
we show that the use of the coda part of the cross-correlation may be not enough to compensate for the seasonal fluctuations when scattering is moderate and an adequate normalization (spectral whitening) of the cross-correlation functions reduces the effect of the seasonal 
fluctuations of the noise sources. We also study the Santorini unrest event of 2011-2012, a slow event that spans over a period of several months,  
and for which it would have been extremely difficult to follow the variations of $dv/v$ without removing the seasonal fluctuations. 
Our results show a decrease in the velocity of seismic waves in the caldera of Santorini which is correlated with the accumulated elevation
measured with GPS. This example illustrates the potential of developing monitoring tools which provide accurate results even with sparse seismic networks with careful signal processing of passive noise recordings.   

\section*{Acknowledgements}
The work of G. Papanicolaou was
partially supported by AFOSR grant FA9550-11-1-0266.  The work of
J. Garnier was partially supported by ERC Advanced Grant Project MULTIMOD-26718.
The work of E. Daskalakis and C. Tsogka was partially supported by the ERC 
Starting Grant Project ADAPTIVES-239959 and the PEFYKA project within the KRIPIS action of the GSRT.

\bibliography{biblio}
\bibliographystyle{plain}%apalike}

\newpage

\appendix

\section{Description of the numerical model.}
In this section we give further details on the numerical model used in section \ref{sec:num}. 
\subsection{The noise sources.}
\label{app:noisesources}%
The function $n(t,\mathbf{x})$ in equation \eqref{eq:1} models the noise sources. We assume that it is a zero-mean random process.
We also assume that the process is stationary in time with 
a covariance function that is delta correlated in space. Therefore, the covariance function of the noise sources has the form
\begin{equation}
\label{eq:2}
\left<n(t_1,\mathbf{y}_1),n(t_2,\mathbf{y}_2)\right>=\Gamma(t_2-t_1,\mathbf{y}_1)\delta(\mathbf{y}_2-\mathbf{y}_1).
\end{equation}
Here $\left<\cdot\right>$ stands for statistical averaging.
The function $t \to \Gamma(t,\mathbf{y})$ is the time correlation function of the noise signals emitted by the noise sources at location $\mathbf{y}$.
The Fourier transform $\omega \to \hat{\Gamma}(\omega,\mathbf{y})$ is their power spectral density (by Wiener-Khintchine theorem).
The function $\mathbf{y} \to \Gamma(0,\mathbf{y})$ characterizes the spatial support of the sources. In our case we assume that the sources are uniformly distributed on a circle ${\cal C}$ of radius $R_{{\cal C}}=25$km as illustrated in Figure \ref{fig:1}:
$$
\Gamma(t,\mathbf{y}) =\frac{1}{2\pi R_{{\cal C}}} \Gamma_0(t,\mathbf{y}) \delta_{{\cal C}}(\by) .
$$
We also assume that we have two receivers at $\mathbf{x}_1=(-5,0)$km and $\mathbf{x}_2=(5,0)$km.

\subsection{Obtaining the time-series data at $\mathbf{x}_1$ and $\mathbf{x}_2$.}
To obtain data at $\mathbf{x}_1$ and $\mathbf{x}_2$ we define the exact distribution and power spectral density of the sources. 
From now on we assume that the statistics of the noise sources change from one day to another and we denote by $\Gamma^j_0(t,\mathbf{y})$ its covariance function at day $j$.
We take  $N_s=180$ point sources uniformly distributed on the circle ${\cal C}$ and then the equation (\ref{eq:4}) becomes
\begin{equation}
\label{eq:6}
\hat{u}^j(\omega,\mathbf{x})=\frac{1}{N_s}  \sum\limits_{i=1}^{N_s} \hat{G}^j(\omega,\mathbf{x},\mathbf{y}_i)\hat{n}_i^{ j}(\omega),
\end{equation}
where $\hat{n}_i^{j}(\omega)$ is the frequency content of the noise sources at $\mathbf{y}_i$ during day $j$, which is random 
such that $\left<  \hat{n}_i^{ j}(\omega)\right>=0$ and 
$$
\left< \hat{n}_i^{ j}(\omega) \overline{\hat{n}_i^{ j}}(\omega')\right> =
2\pi \hat{\Gamma}^j_0(\omega,\by_i) \delta(\omega-\omega') .
$$
At first we consider that the noise sources do not have any seasonal variations and therefore their power spectral density does not depend on $j$. Later on that will be changed according to the model of seasonal variations we want to study. In either case, the last step in order to obtain the time series recorded at location $\mathbf{x}$  is to apply the inverse Fourier transform to (\ref{eq:6}).
\subsection{Relation between the number of days used in the current $\CC$ function and the quality of the measurement obtained by the stretching method} \label{ap:error}
\begin{figure}
\begin{center}
\includegraphics[width=0.45 \textwidth]{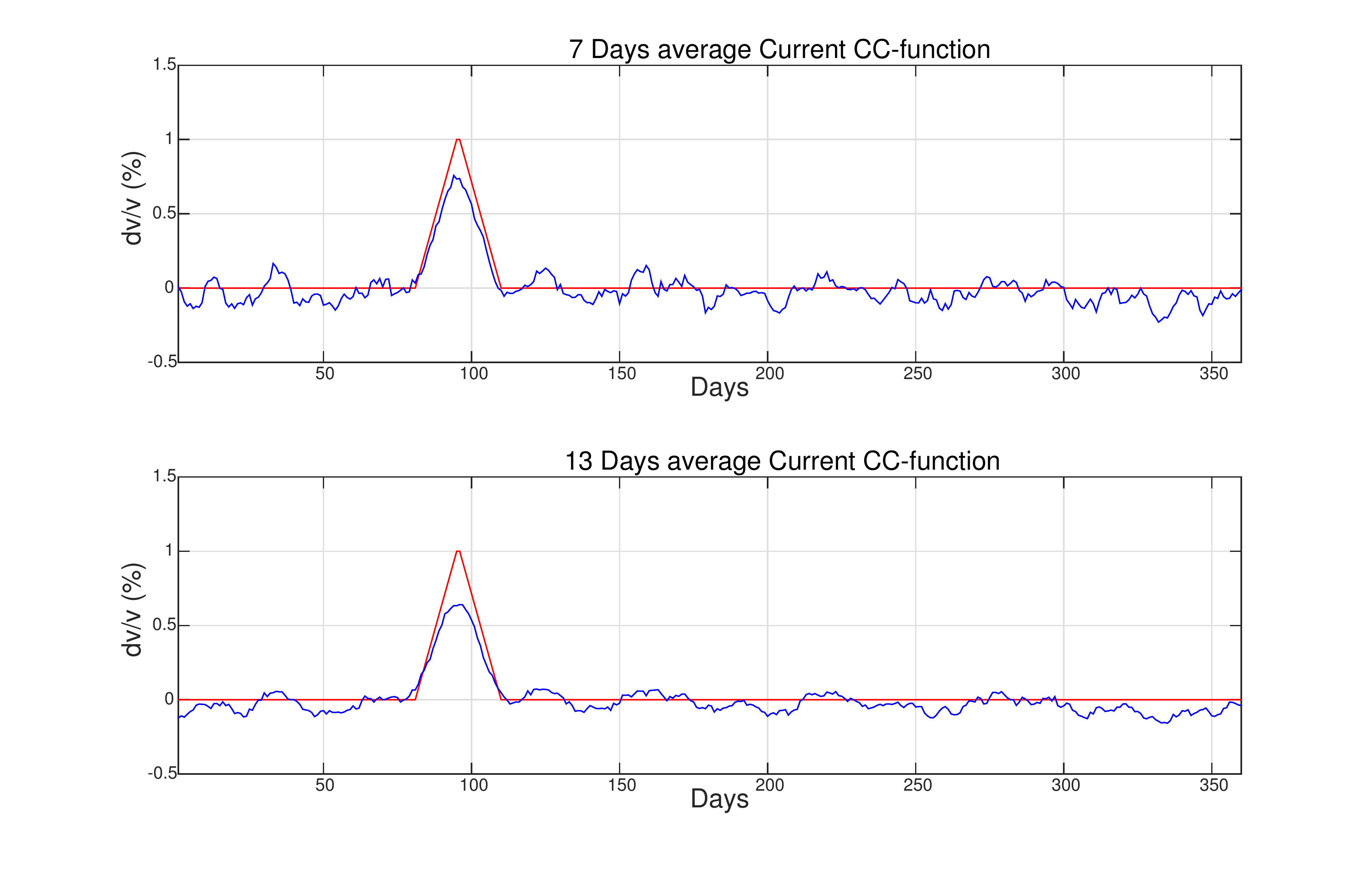}
\caption{In the top plot $N_{ccc}=7$ days are used in computation of the reference $\CC$ while $N_{ccc}=13$ days are used in the bottom plot.
In red is the true velocity variation and in blue the estimated one. 
Using $N_{ccc}=7$ days gives a more precise estimation for the maximal value of $dv/v$ while with $N_{ccc}=13$ days the fluctuations around zero are decreased. }
\label{fig:App_comp}
\end{center}
\end{figure}

\begin{figure}
\begin{center}
\includegraphics[width=0.45 \textwidth]{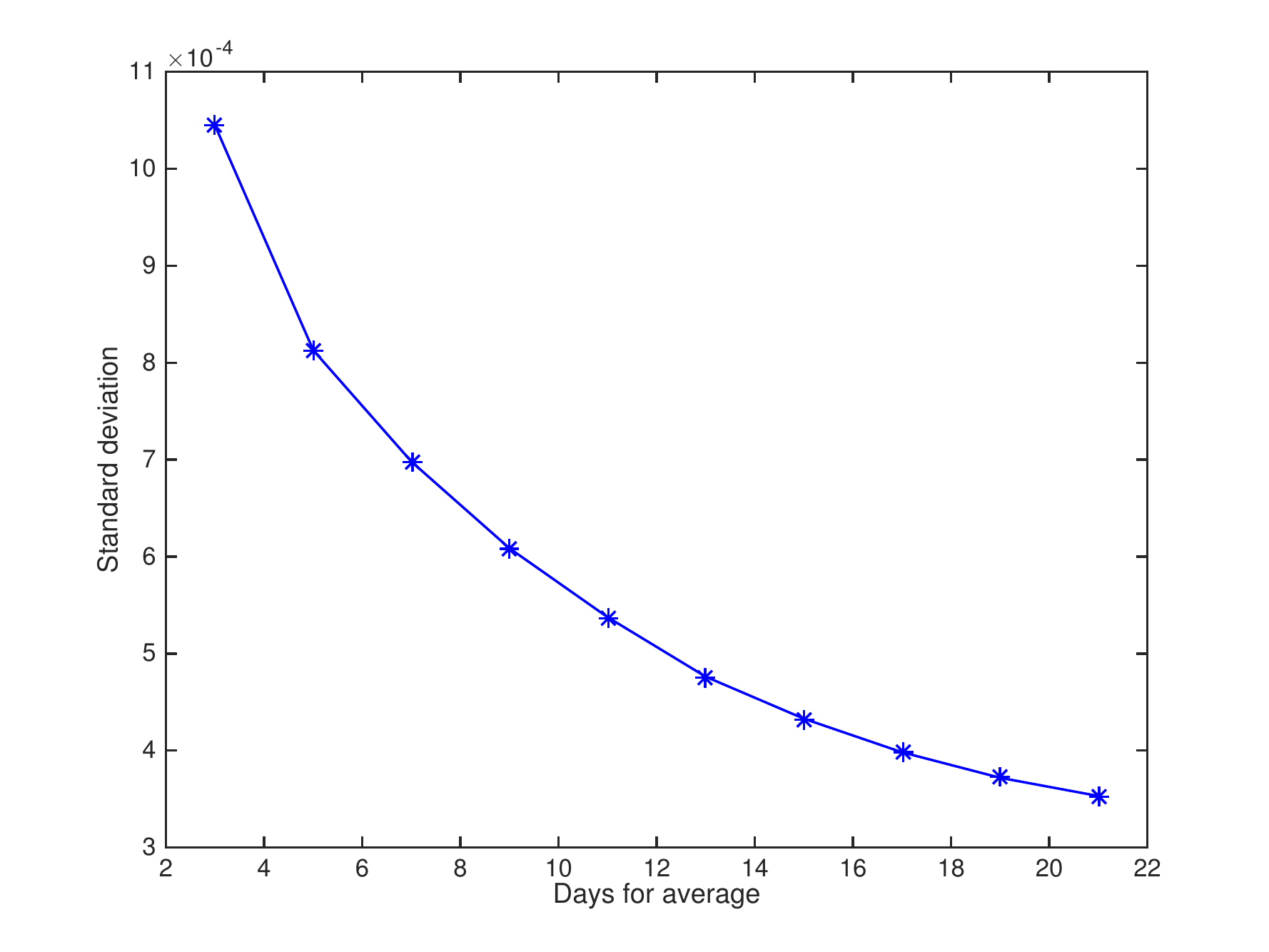}
\includegraphics[width=0.45\textwidth]{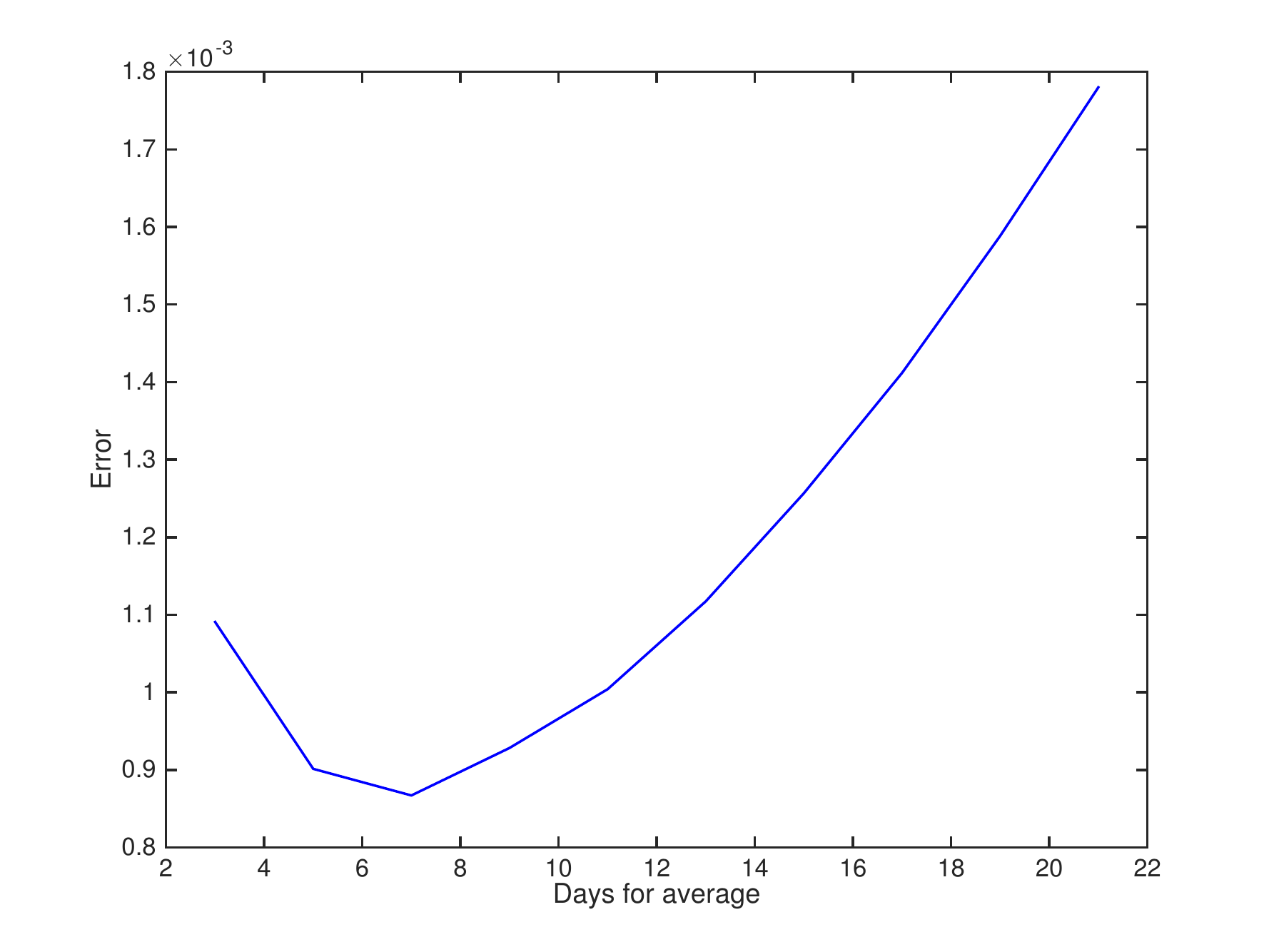}
\caption{Left: The standard deviation of the error in the period where $dv/v=0$ (days 1 to 80 and 110 to 360) as a function of number of days $N_{ccc}$ stacked for the Current \CC function.  
Right: The error for the days 80 to 110 using the norm $\|x\|=\sqrt{\sum\limits_{i=1}^{m}|x_i|^2 }$, where $ x\in\mathbf{R}^m$  as a function of $N_{ccc}$. } 
\label{fig:App_error}
\end{center}
\end{figure}

There is a direct relation between the number of days $N_{ccc}$ that are used in the current $\CC$ function and the standard deviation of the measurement error.  When there is no velocity variations ($dv/v=0\%$),  the obvious answer is that the standard deviation of the error is reduced by increasing the number of days used in the computation of the current $\CC$. However, this results to a loss in precision in the estimation of $dv/v \neq 0$ as illustrated by the results in Figure \ref{fig:App_comp}.
An optimal value for the number of days to be used can be obtained by studying how the error changes as we increase the number of days $N_{ccc}$. 
The value we selected is $7$ since for this value we have a minimum in the error as suggested by the plots in Figure \ref{fig:App_error}, is $N_{ccc}=7$ days. 

\subsection{Uniform and non-uniform seasonal variations.}
\label{app:uniformornot}%
 Our model for the power spectral density of the noise sources is  
 $$
 \hat{\Gamma}_0^j(\omega,\by)=\hat{F}(\omega)  
 \hat{s}^j(\omega,\mathbf{y}),
 $$
 Here the unperturbed noise source distribution is uniform over the circle ${\cal C}$ and has power spectral density $\hat{F}(\omega)$, and 
 $\hat{s}^j(\omega,\mathbf{y})$ is the daily perturbation of the power spectral density at location $\mathbf{y}$. We have two different representations for $\hat{s}^j$:
\begin{enumerate}
\item The daily perturbation is uniform with respect to the locations of the sources: \begin{equation} \label{sjl}
\hat{s}^j(\omega,\mathbf{y})=\hat{f}^j(\omega)l(\mathbf{y}),
\end{equation}
\item The daily perturbation is not uniform and we cannot write it in a separable form. 
\end{enumerate}

In the first case equation (\ref{eq:11}) becomes
\begin{equation}
\label{eq:14}
\begin{array}{l}
\hat{\CC}^j(\omega,\mathbf{x}_1,\mathbf{x}_2)= \hat{F}(\omega)\hat{f}^j(\omega)
\\[7pt]
\hspace*{0.0cm}\dsp  \times \int_{\cal C} d\sigma(\by) 
\; \overline{\hat{G}^j(\omega,\mathbf{x}_1,\mathbf{y})}\hat{G}^j(\omega,\mathbf{x}_2,\mathbf{y})l(\mathbf{y}) ,
\end{array}
\end{equation}
and it is clear that after spectral whitening, any daily perturbation in the power spectral density of the noise sources will be eliminated since the perturbation is contained into the amplitude spectra of the cross-correlation function. 
In the second case we cannot  separate the terms due to the sources and take them out of the integral. \\
Instead of equation (\ref{eq:6}), we use,
\begin{equation}
\label{eq:15}
\begin{array}{l}
\dsp \hat{u}^j(\omega,\mathbf{x})=\frac{1}{N_s} \sum\limits_{i=1}^{N_s} \hat{n}^j_i(\omega)\hat{G}^j(\omega,\mathbf{x},\mathbf{y}_i)  
\\[7pt]
%\hspace*{2.3cm}
 \times (1-\delta \hat{g}(\omega)\sin(2\pi j /N_d)),
\end{array}
\end{equation}
with $\delta=0.4$ and 
\[   
\hat{g}(\omega) = 
     \begin{cases}
       1 & \text{if}\;\;\omega_1\leq \omega\leq\omega_1+ \pi B,\\
      0 &\text{if}\;\; \omega_1+\pi B<\omega\leq \omega_1+ 2\pi B,\
     \end{cases}
\]
to simulate uniform seasonal variations with
\begin{equation}
\label{sj1}
\hat{s}^j(\omega,\mathbf{y}) = (1-\delta \hat{g}(\omega)\sin(2\pi j /N_d))^2.
\end{equation}
In the simulations we take $\hat{F}(\omega)={\bf 1}_{[\omega_1,\omega_1+2\pi B]}(|\omega|)$,
 $B=0.5$Hz and $\omega_1= 2 \pi \, 0.15$rad.s${}^{-1}$.
To add anisotropy we multiply \eqref{eq:15} by a function that depends on the source azimuth, $\theta(\by)$. More precisely, we take
\begin{equation}
\label{eq:15b}
\begin{array}{l}
\dsp \hat{u}^j(\omega,\mathbf{x})=\frac{1}{N_s} \sum\limits_{i=1}^{N_s} \hat{n}^j_i(\omega)\hat{G}^j(\omega,\mathbf{x},\mathbf{y}_i)  
\\[7pt]
%\hspace*{2.3cm}
 \times (1-\delta \hat{g}(\omega)\sin(2\pi j /N_d)) (1-0.6 \cos{(2\theta(\by_i))}),
\end{array}
\end{equation}
which results to a model for $\hat{s}^j(\omega,\by)$ in \eqref{eq:14} of the form 
\begin{equation}
\label{sjanis}
\hat{s}^j(\omega,\by) = (1-\delta \hat{g}(\omega; \theta(\mathbf{y})+2\pi j / N_d)\sin(2\pi j /N_d))^2 (1-0.6\cos{(2\theta(\by))})^2,
\end{equation}
where $\theta(\mathbf{y})$ is the angle of $\mathbf{y}$ on the circle ${\cal C}$.
This is a quite extreme case of anisotropy cf. \cite{weaver09,colombi} which allows us to illustrate the robustness of the proposed filtering.
For the non-uniform case, we use,
\begin{equation}
\label{eq:16}
\begin{array}{l}
\dsp \hat{u}^j(\omega,\mathbf{x})= \frac{1}{N_s}
\sum\limits_{i=1}^{N_s} \hat{n}^j_i(\omega)\hat{G}^j(\omega,\mathbf{x},\mathbf{y}_i)
\\[5pt]
\hspace*{0.0cm}
 \times (1-\delta \hat{g}(\omega;2\pi i / N_s+2\pi j / N_d)\sin(2\pi j /N_d)),
\end{array}
\end{equation}
where
\begin{equation}
\label{eq:16b}
\hat{g} (\omega;\theta ) = 
     \begin{cases}
       1 & \text{if}\;\;\omega_1\leq \omega\leq \omega(\theta ) ,\\
      0 &\text{if}\;\; \omega(\theta )<\omega\leq \omega_1+ 2\pi B, \
     \end{cases}
\end{equation}
with
\begin{equation}
\label{eq:17}
\omega(\theta )=\omega_1+\pi B + \pi B\sin(\theta ).
\end{equation}
This models non-uniform seasonal variations with
\begin{equation}
\label{sj2}
\hat{s}^j(\omega,\by) = (1-\delta \hat{g}(\omega; \theta(\mathbf{y})+2\pi j / N_d)\sin(2\pi j /N_d))^2.
\end{equation}

\label{lastpage}

\end{document}